\documentclass[aps,pre,longbibliography,twocolumn,floatfix,superscriptaddress]{revtex4-1}
\usepackage{amsmath,amssymb,graphicx,cases}
\usepackage{epsfig}
\usepackage{textcomp}
\usepackage{epstopdf}
\usepackage{float}
\usepackage{braket}
\usepackage{comment}
\usepackage{tabularx}
\usepackage[normalem]{ulem}

\usepackage[normalem]{ulem}
\newcommand{\stkout}[1]{\ifmmode\text{\sout{\ensuremath{#1}}}\else\sout{#1}\fi}

\usepackage{color}
\usepackage[dvipsnames]{xcolor}
\definecolor{Blue}{rgb}{0.00, 0.00, 1.00}
\definecolor{Red}{rgb}{1.00, 0.00, 0.00}
\definecolor{Green}{rgb}{0.00, 0.50, 0.00}

\newcommand{\be}{\begin{equation}}
\newcommand{\ee}{\end{equation}}
\newcommand{\bea}{\begin{eqnarray}}
\newcommand{\eea}{\end{eqnarray}}

\usepackage[colorlinks=true, urlcolor=blue, anchorcolor=blue, citecolor=blue,filecolor=blue,linkcolor=blue,menucolor=blue]{hyperref}

\usepackage{fixmath}

\makeatletter
\def\@bibdataout@aps{%
\immediate\write\@bibdataout{%
@CONTROL{%
apsrev41Control%
\longbibliography@sw{%
    ,author="08",editor="1",pages="1",title="0",year="1"%
    }{%
    ,author="08",editor="1",pages="1",title="",year="1"%
    }%
  }%
}%
\if@filesw \immediate \write \@auxout {\string \citation {apsrev41Control}}\fi 
}
\makeatother

\begin{document}
\title{Data-driven analysis of annual rain distributions}

\author{Yosef Ashkenazy}
\email{ashkena@bgu.ac.il}
\affiliation{Department of Environmental Physics, Blaustein Institutes for Desert Research, Ben-Gurion University of the Negev, Sede Boqer Campus, 8499000, Israel}
\author{Naftali R. Smith}
\email{naftalismith@gmail.com}
\affiliation{Department of Environmental Physics, Blaustein Institutes for Desert Research, Ben-Gurion University of the Negev, Sede Boqer Campus, 8499000, Israel}


\begin{abstract}
Rainfall is an important component of the climate system and its statistical properties are vital for prediction purposes. In this study, we have developed a statistical method for constructing the distribution of annual precipitation. The method is based on the convolution of the measured monthly rainfall distributions and does not depend on any presumed annual rainfall distribution. Using a simple statistical model, we demonstrate that our approach allows for a better prediction of extremely dry or wet years with a recurrence time several times longer than the original time series. The method that has been proposed can be utilized for other climate variables as well.
\end{abstract}

\maketitle

\section{Introduction}
Terrestrial precipitation is a crucial component for the survival of humans and has a significant impact, e.g., on water supplies, river flow, and dryland farming. Additionally, precipitation has a significant impact on water usage in irrigated agriculture, industry, and hydroelectric power generation. It is thus important to be able to make reliable quantitative predictions of rain distributions, to design and manage irrigation systems, etc. One may be particularly interested in estimating the likelihood of rare events such as droughts or floods, which can have devastating effects. These problems are becoming of growing importance, and in particular, water scarcity is a significant problem faced by many societies today \cite{Pereira-Cordery-Iacovides-2009:coping}. According to the recent sixth IPCC (Intergovernmental Panel on Climate Change) report, about half the world population faces severe water scarcity each year due to climate and other factors \cite{Pörtner-Roberts-Adams-et-al-2022:climate} and humans are contributing more to extreme weather, especially with regard to drought and precipitation \cite{Masson-Delmotte-Zhai-Pirani-et-al-2021:climate}. It was also predicted that global warming will lead to increased heavy precipitation and worsened droughts in certain regions \cite{Masson-Delmotte-Zhai-Pirani-et-al-2021:climate}. Throughout history, civilizations have collapsed and settlements have shifted due to extreme changes in precipitation \cite{Weiss-Courty-Wetterstrom-et-al-1993:genesis, Berkelhammer-Sinha-Stott-et-al-2012:abrupt, Kuper-Kropelin-2006:climate, Faust-Ashkenazy-2007:excess, Faust-Ashkenazy-2009:settlement, Griffiths-Johnson-Pausata-et-al-2020:end}.

Accurately understanding the accumulated precipitation statistics over extended periods such as annual rainfall, as focused on in this work, is crucial, e.g., for efficient water management and crop planning. Typically, forecasts are based on a statistical analysis of the precipitation time series \cite{Brown-Katz-Murphy-1986:economic}. This approach is expected to work reasonably well because the annual rain statistics appear relatively robust to current climate change in some regions. For instance, the annual rain over most of the global land surface was found to be fairly stationary over the last century \cite{Sun-Roderick-Farquhar-2018:rainfall, Ukkola-Roderick-Barker-et-al-2019:exploring}, although over shorter timescales, extreme weather events are becoming likelier \cite{Donat-Lowry-Alexander-et-al-2016:more, Masson-Delmotte-Zhai-Pirani-et-al-2021:climate}. 

With regards to precipitation in Israel which is investigated in this paper, one study showed a change over the period 1930-1990 in the shape parameters of fitted rain distributions in Israel \cite{Ben-Gai-Bitan-Manes-et-al-1998:spatial}, although a subsequent study showed no significant trend in Israel \cite{Alpert-Ben-Gai-Baharad-et-al-2002:paradoxical}. A third study showed mixed results \cite{Kafle-Bruins-2009:climatic}. Yet another study showed a shortening of the rainy season in Israel \cite{Ziv-Saaroni-Pargament-et-al-2014:trends}. 
Our analysis (not shown) indicates a drop in annual precipitation in Israel, between 1980 and 2019 (including) where only 5 stations out of 110 exhibited increasing trends. The reason for the above, somehow contradicting, studies may be rooted in the fact that the different studies analyzed different periods, capturing different parts of much slower decadal and centennial cycles. 

\begin{figure*}[t]
\includegraphics[width=0.8\linewidth]{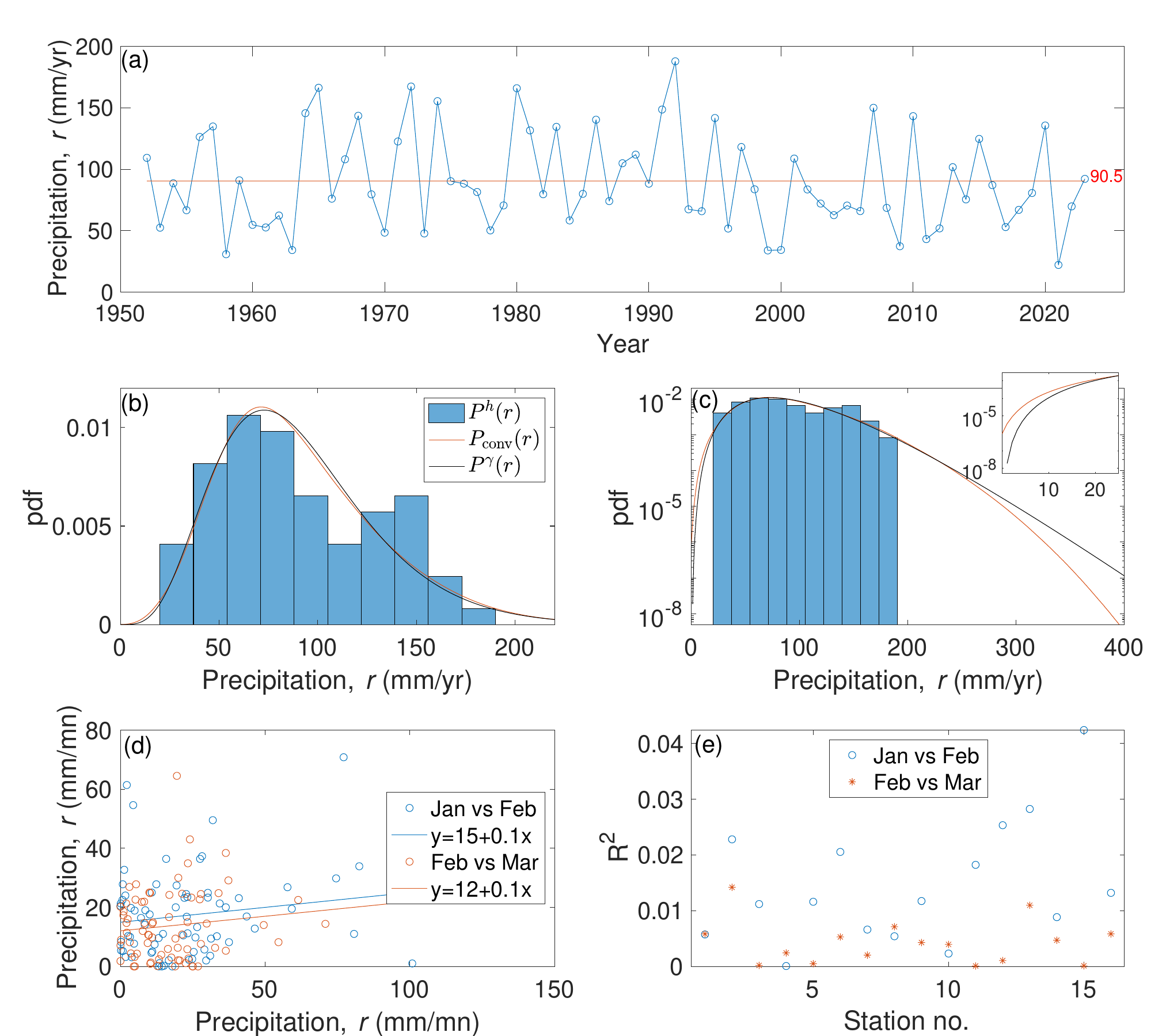}
\caption{(a) Annual rainfall (Sep. to August) versus time (year) in Sede Boker, for the period 1952-2023, and the corresponding histogram, $P^h(r)$, in (b) regular and (c) semi-log plots. In (b) and (c), the red line indicates the PDF, $P_{\text{conv}}(r)$, constructed based on the method reported in this paper, and the black line represents a fit to a gamma distribution, $P^\gamma(r)$. The inset in (c) depicts an enlargement of the left side of the distribution. (d) Feb. precipitation versus Jan. precipitation (blue circles), Mar. precipitation versus Feb. precipitation (red circles), and their corresponding linear fits (blue and red solid line). Note the absence of correlation between the rains of different months. (e) The $R^2$ of Feb. rain (over the different years) versus Jan. rains (empty blue circles) and March rain versus Feb. rain (full red circles) of 16 stations scattered over Israel (see details in Supplementary Table \ref{tab:station_details}). Note the very small $R^2$, indicating the absence of correlations between the different months. 
}
\label{figAnnualRainSB}
\end{figure*}

Predicting the likelihood of atypical annual rainfall events, like extreme droughts, has been a challenge for existing forecasts. This is because such events have occurred only a few times in the historical data or not at all. One could attempt to do this by fitting the data to some standard distribution \cite{Yue-Hashino-2007:probability, Paulo-Martins-Pereira-2016:influence} (often, gamma distributions are used), and then extrapolating the fitted distribution to its tails. However, this would be problematic for two reasons, which are both related to the fact that reliable historical rainfall data exists for the last century or two at most. 
First, different studies used different distributions to fit the annual rain data where the choice was arbitrary to some degree; the goodness of fit may be of similar level when the number of observations (years) rarely exceeds 100; see for example Fig.~\ref{figAnnualRainSB} and Fig. \ref{fig_stations_time_series}.
Second, the fit is performed on the main part of the distribution, describing typical fluctuations.  However, the fit is expected to break down in the distribution tails, as for a wide variety of statistical models in which the distributions display very different behaviors in the central part and the tails \cite{Varadhan-1984:large,Oono-1989:large,Dembo-Zeitouni-2009:large,Hollander-2000:large,Majumdar-Schehr-2017:large}.

One alternative approach would be to use climate models to make the predictions \cite{Ragone-Wouters-Bouchet-2018:computation, Ragone-Bouchet-2021:rare, Galfi-Lucarini-Ragone-et-al-2021:applications, Galfi-Lucarini-2021:fingerprinting, Errico-Pons-Yiou-et-al-2022:present}. This approach has some merit; however, one may still be concerned about how well the models correctly describe the real system, especially in its large-deviation regime. Moreover, high-resolution models (e.g., atmospheric General Circulation Models, GCMs) are computationally expensive to run for extended periods of simulation time, which is necessary for predicting rare events (this difficulty can be somewhat mitigated by using large deviations simulation algorithms \cite{Ragone-Wouters-Bouchet-2018:computation, Ragone-Bouchet-2021:rare, Galfi-Lucarini-Ragone-et-al-2021:applications, Galfi-Lucarini-2021:fingerprinting, Errico-Pons-Yiou-et-al-2022:present}).
Another approach is to combine data from many different locations. This increases the amount of data considerably but involves assuming that different locations behave similarly. For instance, one can assume that, up to scaling factors, the annual rainfall in different locations follows the same distribution. This is, however, not the case for annual precipitation in Israel since the spatial variability is relatively large--see Supplementary Table \ref{tab:station_details}.

In this paper, we develop an alternative method for predicting annual rainfall, which is based solely on measured data, without a need for performing any fits on observed distributions (using one of many distributions that were used to fit rain data). Our method does not involve any parameters that need to be fitted to observed data either. The method exploits the timescale separation between the correlation time of weather (synoptic) events---typically on the order of one week---and the timescale of one year over which the rain accumulates. It enables us to make quantitative predictions for annual rainfall statistics, outside the central part of the distribution.
In particular, it enables us to predict the likelihood of extremely dry or wet years beyond observed data.

While we focus on annual rain distributions in Israel, 
our method is quite versatile, and can be used for any geographical region, and for any quantity that is accumulated/averaged over a period of time much longer than the typical correlation time of the dynamical system in question. The method is based on dividing the time series under consideration into time windows of intermediate duration, which are approximated to be uncorrelated. Some modifications may needed to apply the method, e.g., on seasonally averaged temperatures \cite{Ciavarella-Cotterill-Stott-et-al-2021:prolonged, Galfi-Lucarini-Ragone-et-al-2021:applications, Galfi-Lucarini-2021:fingerprinting}.

\section{Annual rainfall distribution}
Let us consider, as a particular example, the annual rainfall in Sede Boker in Israel, for the period 1952-2022; see Fig.~\ref{figAnnualRainSB}a. The data of this station as well as the data of 15 additional stations are all downloaded from the Israeli Meteorological Service web page \href{https://ims.gov.il/en/data_gov}{https://ims.gov.il/en/data\_gov}; see Supplementary Table \ref{tab:station_details} and Fig. \ref{fig_stations_time_series}. Sede Boker is located in the Negev desert, in an extreme arid region with an average annual rainfall of $\approx$90 mm/year. The annual rainfall time series does not appear to follow a clear increasing or decreasing temporal trend; this is also the situation for the other 15 stations' data we studied. Following this observation and Hasselmann \cite{Hasselmann-1976:stochastic} we have calculated the correlation coefficient $R^2$ of the annual precipitation time series (i.e., 1-year lag cross-correlation) and found no significant correlations in all 16 stations under consideration. Moreover, the rainfall of consecutive months also appears uncorrelated (Fig.~\ref{figAnnualRainSB}e,f). We thus approximate the rainfall of the different months as independent random variables. 
The goal of this work is to estimate, based on the data, the PDF $P(r)$ that describes the statistics of the annual rainfall $r$ in a given location (e.g., Sede Boker). 

A simple estimate $P^h(r)$ for the annual rain PDF is obtained by constructing a histogram of the annual rainfall distribution; see Fig.~\ref{figAnnualRainSB}b,c. However, since this histogram is not based on a very large number of years (typically less than 100 years), it does not give a good estimate of the $P(r)$; see the bi-modal distribution presented in Fig.~\ref{figAnnualRainSB}b (and the PDFs of the other stations analyzed here, Fig.~\ref{fig_stations_pdf}) while the PDF is expected to be uni-modal. As mentioned in the Introduction, a common practice is to fit $P^h(r)$ to a known distribution (like the gamma or some other standard distributions); such a fit (i.e., fit to the gamma distribution) is plotted in Fig.~\ref{figAnnualRainSB}b,c. One can then treat this fit, $P^\gamma(r)$, as a prediction for the true PDF $P(r)$. However, the choice of the fitted distribution is in some sense arbitrary and may lead to different predictions, especially of extreme precipitation years. In addition, it is difficult to assess the reliability of such predictions. This difficulty becomes especially pronounced in the tails of the distribution, where the probabilities are very small, making it impossible to quantitatively test the accuracy of $P^\gamma(r)$ due to insufficient data.


\subsection{The convolution PDF method}

The key idea behind the method we propose is to use additional information, beyond the total annual rainfall data; an example of annual rain time series is plotted in Fig.~\ref{figAnnualRainSB}a and Fig. \ref{fig_stations_time_series}. Indeed, rainfall data is often recorded at much higher resolutions (like 10 minutes), and as we show below, one can make use of this additional information to predict the annual rain PDF to much higher accuracy, including, to a certain extent, the prediction of extreme annual precipitation corresponding to the tails of the distribution.

We exploit the separation of scales between the typical timescale $\tau$ for correlations in weather data (typically, of the order of one week) and the much larger timescale of an entire year. We do this by dividing the year into intermediate units of time that are much shorter than a year but still much longer than $\tau$. A convenient choice we adopt here is to take these time units to be the 12 months of the year, but in principle, we would expect any choice between $\sim 15$ and $\sim 60$ days to work reasonably well. (One advantage of dividing the data into months is that some historical rainfall data is given at a monthly resolution.) Our method can be summarized as follows: We write the annual rainfall as the sum $r = r_1 + \dots + r_{12}$ of the monthly rainfalls $r_i$, where $r_i$ are regarded as statistically independent.
We tested our approximation of absence of correlation between the monthly rains by calculating the $R^2$ of the Feb. versus Jan. rains and Mar. versus Feb. rains (Fig.~\ref{figAnnualRainSB}d,e); the $R^2$ of all 16 stations under consideration (Supplementary Table \ref{tab:station_details}) is very low, indicating the absence of correlations, validating our approximation. We note that precipitation in Israel falls during the winter months (Nov. to May) and is nearly absent during the summer months; see Supplementary Fig. \ref{figMontlhyRain}.
%

The next step in the analysis is to construct monthly rain histograms 
$P_{1}^{h}(r_{1}),\dots,P_{12}^{h}(r_{12})$ corresponding to each of monthly rainfall time series $r_1, \dots, r_{12}$. We use these histograms as approximations of the true monthly PDFs $P_{1}(r_{1}),\dots,P_{12}(r_{12})$. Based on the approximation of statistical independence between the rainfall of the different months (Fig.~\ref{figAnnualRainSB}d,e), our estimate for the annual rainfall PDF $P(r)$ is given by the convolution of the 12 monthly histograms, i.e.,
\be
\label{PconvDef}
P_{\text{conv}} \! \left(r\right) \! = \!\! \int \!\! dr_{1}\cdots \! \int \!\!dr_{12}P_{1}^{h}\left(r_{1}\right)\cdots P_{12}^{h}\left(r_{12}\right)\delta \! \left(\! r \! - \! \sum_{i=1}^{12}r_{i} \! \right) \! .
\ee
Before testing the quantitative predictions of our method, let us briefly discuss its expected advantages and disadvantages. First, we do not assume an underlying model or theoretical distribution. 
Our only assumptions are (as discussed above) that, to a good approximation, the system's statistics do not change from year to year and that $r_1, \dots, r_{12}$ are statistically independent (Fig.~\ref{figAnnualRainSB}d,e). Under these assumptions, one expects our method to yield a more accurate and reliable prediction than the rough estimate of the annual rain histogram $P^{h}\left(r\right)$ (bars in Fig.~\ref{figAnnualRainSB}b,c) that is based on annual rainfall data. In particular, our method yields predictions of the annual rain PDF for a wider range of $r$'s, including in the distribution tails.

We now conduct a quantitative analysis of our results for data from the Sede Boker station and 15 other stations (Supplementary Table \ref{tab:station_details}). Moreover, to validate the proposed method, we test its predictions on a (very) simplified statistical model of annual rainfall.

\subsection{Results}

\begin{figure*}[t]
\includegraphics[width=0.9\linewidth,clip=]{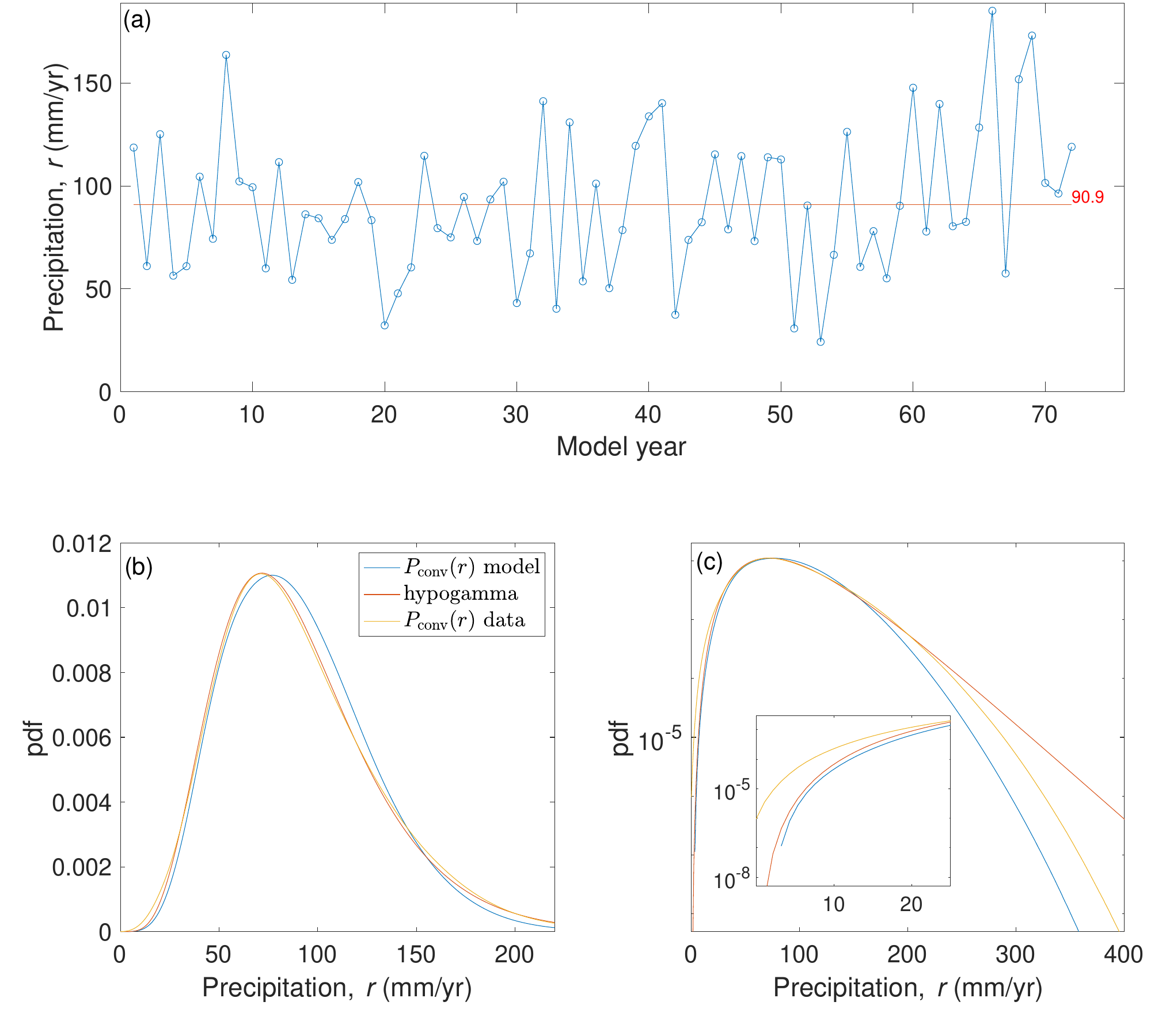}
\caption{(a) A typical annual rainfall time series generated by our toy model with parameters based on Sede Boker data shown in Fig. \ref{figAnnualRainSB}. (b) The PDF calculated by our convolution method, $P_{\rm conv}(r)$ (blue line) using the time series shown in (a). The theoretical hypogamma distribution \cite{Moschopoulos-1985:distribution} (red line) of the suggested model that was used to generate the time series shown in (a). For comparison, we also include the convolution PDF based on the actual Sede Boqer annual rainfall data (also shown in Fig. \ref{figAnnualRainSB}b). (c) Same as (b) in semi-log representation. The inset shows an enlargement of the left side of the distribution. }
\label{figResultsModel}
\end{figure*}

Our prediction $P_{\text{conv}}\left(r\right)$ for the PDF of the annual rainfall in Sede Boker is shown in Fig.~\ref{figAnnualRainSB}b,c (red line). We note that the predicted PDF is a smooth curve, although no fitting to a smooth theoretical curve was performed. Moreover, unlike the histogram $P^h(r)$ of the annual rainfall data (blue bars) which is bi-modal, the PDF calculated by our convolution method is uni-modal; this is the situation for some other stations analyzed in this study (Fig.~\ref{fig_stations_pdf}). In the regime of typical fluctuations, $r=30-200$ mm/year, our prediction is close to the fitted gamma distribution $P^\gamma(r)$ (black line), as seen in Fig.~\ref{figAnnualRainSB}b,c. However, in the left ($r\lesssim 10$ mm/year) and right ($r \gtrsim 250$ mm/year) tails of the distribution, the two predictions differ by orders of magnitude: our predicted PDF is much larger (smaller) than the gamma distribution in the left (right) tail. 
The inset in Fig.~\ref{figAnnualRainSB}c indicates that the left tail of the gamma distribution, $P^\gamma(r)$ lays below the the convolution distribution, $P_{\rm conv}(r)$; this is expected as the probability for zero annual rainfall for the extreme arid Sde-Boqer station is finite while the probability density of the gamma PDF is by definition zero.

 To estimate the size of the errors in our prediction $P_\text{conv}(r)$, we also performed bootstrapping as follows. We generated a large number of realizations of $N=72$ years, where each year was created by randomly sampling each of the monthly precipitations $r_i$ from the data.
For each realization, we then created the histogram $P^h(r)$. The results of the bootstrapping are plotted in Fig.~\ref{fig:bootstrapping}. As is seen in the figure, in the central part of the distribution, the errors are of order $10\%$, while in the tails they become much larger.

There is no reason to believe that the gamma distribution, or other theoretical distributions that are used to fit the annual rain distribution, captures the distribution tails correctly. Generically, in large deviation theory, one finds that theoretical curves that correctly capture the regime of typical fluctuations do not faithfully describe the tails of the distribution \cite{Varadhan-1984:large,Oono-1989:large,Dembo-Zeitouni-2009:large,Hollander-2000:large,Majumdar-Schehr-2017:large}. For example, the distribution of a particle undergoing a random walk, and the distribution of the largest eigenvalue of a Gaussian unitary ensemble (GUE) random matrix are described by the Gaussian and Tracy-Widom distributions, respectively, in the typical fluctuations regimes, but these predictions break down in the tails of the distributions \cite{Majumdar-Schehr-2014:top, Majumdar-Schehr-2017:large}.


\subsection{Rate of convergence of our method}

The accuracy of our method is expected to improve as more data is accumulated. However, for any finite amount of data, there will be discrepancies between the predicted PDF and the true annual rain PDF. Generally speaking, we expect relatively larger discrepancies in the tails of the annual rain distribution. 
Still, the two tails of the annual rain distribution may have different qualitative behaviors as the left tail is bounded by zero precipitation $r=0$, making its prediction, most probably, more accurate, as we explain below. 


We expect the predicted annual PDF $P_{\text{conv}}(r)$ to underestimate the true PDF $P(r)$ in the right tail. One way to see this is to notice that $P_{\text{conv}}(r)$ has a finite support: It is truncated at some maximal value that equals the sum of the recorded maximum monthly values. However, there must be some nonzero probability for $r$ to exceed this value---for instance, for a longer time series, this maximum value is expected to be higher.
Moreover, an unusually large total annual rain may involve just \emph{one} of $r_1, \dots, r_{12}$ being much larger than expected. Indeed, if the monthly PDFs $P_{i}\left(r_{i}\right)$ decay exponentially or slower at $r_i \to \infty$, then the $r\to\infty$ tail is expected to be dominated by this scenario.
This situation is called the ``big-jump principle'',  in analogy with discrete-time random walks, in which the tail of the position distribution of the particle after many steps is dominated by scenarios in which a single step contributes more than the sum of all the others. The big-jump principle is known to occur when considering the distributions of sums of random variables (and in analogous situations involving continuous-time dynamical systems) in many physical systems and mathematical models \cite{Chistyakov-1964:theorem, Denisov-Dieker-Shneer-2008:large, Geluk-Tang-2009:asymptotic, Foss-Korshunov-Zachary-et-al-2011:introduction, Vezzani-Barkai-Burioni-2019:single, Wang-Vezzani-Burioni-et-al-2019:transport, Vezzani-Barkai-Burioni-2020:rare, Nickelsen-Touchette-2018:anomalous, Gradenigo-Majumdar-2019:first, Meerson-2019:anomalous, Jack-Harris-2020:giant, Brosset-Klein-Lagnoux-et-al-2020:probabilistic, Gradenigo-Iubini-Livi-et-al-2021:localization, Gradenigo-Iubini-Livi-et-al-2021:condensation, Mori-Le-Majumdar-et-al-2021:condensation, Mori-Gradenigo-Majumdar-2021:first, Smith-2022:anomalous, Smith-Majumdar-2022:condensation}. 
This leads to the (conservative) estimate that, in the right tail, our predictions are reliable up to events whose recurrence time is approximately the number of years of data; however, as shown below, it appears that our predictions are reliable well beyond this conservative estimate. In any case, our prediction gives a lower bound for the true PDF in the right tail.

In contrast, in the left tail, the ``big jump principle'' cannot possibly hold, since the distribution is bounded from below by $r=0$. An unusually small annual rain $r$ must therefore be the result of all (or most) of $r_1,\dots,r_{12}$ being below average. Such a combination of slightly unusual events (namely, small monthly rainfalls), which all may occur within the observed data, may thus enable us to accurately predict the likelihood of an extremely rare event (small annual rainfall).

We quantitatively tested the validity of our method by applying it to a simple toy model. In this model, the monthly rainfall distributions $P_{1}(r_{1}),\dots,P_{12}(r_{12})$ are assumed to be statistically-independent gamma distributions, whose parameters were estimated through the Maximum Likelihood Estimation (MLE), using the measured data of each station. The sum of independent gamma variables, the ``hypogamma'' PDF, $P(r)$, is given in \cite{Moschopoulos-1985:distribution}. We then simulate $N$ years of data and apply our method to the simulated time series. More specifically, we create monthly histograms 
$P_{1}^{h}(r_{1}),\dots,P_{12}^{h}(r_{12})$ and convolute them to obtain the estimate $P_{\text{conv}}(r)$. We then compare this estimated PDF to the exact PDF, $P(r)$. A typical model time series generated using parameters that are based on Sede Boqer data (Fig.~\ref{figAnnualRainSB}a) (and of the same duration, $N=72$ years) is shown in Fig.~\ref{figResultsModel}a. In Fig. \ref{figResultsModel}b,c we depict the corresponding convolution PDF of the model time series shown in Fig.~\ref{figResultsModel}a that was constructed using our method (blue line). We also show the theoretical hypogamma PDF (red line) and the convolution PDF of the original Sede Boqer time series (Fig.~\ref{figAnnualRainSB}b,c). It is apparent that the convolution PDFs of the model and the data are similar.
It is also clear from Fig.~\ref{figResultsModel}c, as suggested above, that the convolution PDF $P_{\text{conv}}(r)$ (blue) reliably predicts the theoretical PDF $P(r)$ (red) outside the range of the model data shown in Fig.~\ref{figResultsModel}a. As expected, $P_{\text{conv}}(r)$ underestimates $P(r)$ sufficiently far into the right tail. In the left tail, the prediction $P_{\text{conv}}(r)$ is very accurate and only breaks down at $r$ which is of the order of a few mm/year (corresponding to extremely rare events).

\begin{figure*}[t]
\includegraphics[width=1\linewidth]{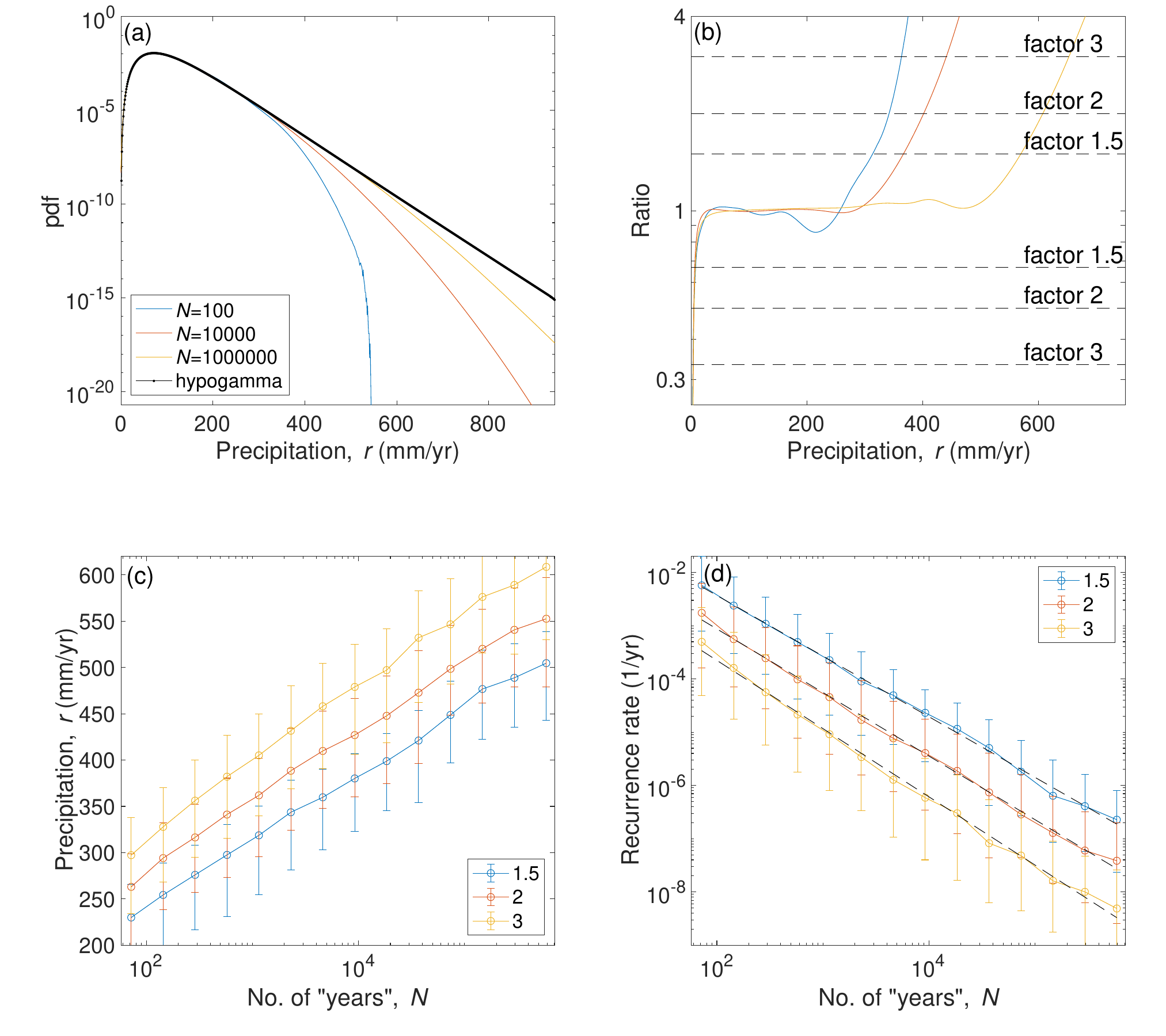}
\caption{
(a) The predicted PDF for model data using an increasing number of data points (years, see legend). The theoretical hypogamma PDF is indicated by the solid black line. (b) The ratio between the theoretical hypogamma PDF $P(r)$ and the convolution PDFs $P_{\text{conv}}(r)$, based on different length time series (see legend of panel a). Note that the ratio is larger (smaller) than one at the right (left) side of the graph, corresponding to the right (left) tail of the distributions shown in (a). (c) The annual precipitation value $r_+$ at which the ratio $P(r)/P_{\text{conv}}(r)$ or $P_{\text{conv}}(r)/P(r)$ exceeds 1.5 (blue), 2 (red), and 3 (yellow) versus the length of the time series (in model years). The results are based on 1000 model realizations (for each $N$) where the errorbars indicate the 25\%-75\% quantile range and the circles indicate the median. (d) Same as (c) for the recurrence rate (in 1/yr) corresponding to $r_+$. Note the power law decay of the graphs with exponents of -1.1, -1.2, and -1.3 (black dashed lines) for the factors 1.5, 2, and 3 respectively.}
\label{figModelRec}
\end{figure*}

We next systematically analyzed how the length $N$ of the annual precipitation time series affects the accuracy of predicted PDFs. In Fig.~\ref{figModelRec}a, we present a few examples of the convolution PDFs, based on time series of different lengths that were generated by the model. For comparison, we also depict the theoretical hypogamma distribution. As expected, the predicted PDF converges to the theoretical one as the number of years $N$ of the model's time series increases. Fig.~\ref{figModelRec}b depicts the ratio between the theoretical hypogamma distribution $P(r)$ and the convolution PDFs $P_{\text{conv}}(r)$ obtained from the model's time series as a function of the precipitation rate $r$. As expected, longer time series exhibit a better match with the theoretical distribution. It is also noticeable that the ratio is larger (smaller) than one on the right (left) side of the graph, i.e., the predicted PDF underestimates (overestimates) the theoretical distribution at the right (left) tail. 
To remind the reader, the right tail of the distribution provides an estimation for the recurrence time of very wet years while the left tail provides an estimation for the recurrence time of very dry years. 

Let us describe first the analysis of the right tail of the distribution. In Fig.~\ref{figModelRec}c,d we study the maximum annual precipitation $r_+$ that can be predicted by our method. We arbitrarily chose a factor of 1.5, 2, and 3-time difference (horizontal dashed lines in Fig.~\ref{figModelRec}b) as the limit of validity; i.e., the ratio between the theoretical and calculated PDFs, $P(r)/P_{\text{conv}}(r)$ or $P_{\text{conv}}(r)/P(r)$ equal to a specific value. To capture these two cases, we studied $|\ln{P_{\text{conv}}(r)/P(r)}|$. Fig.~\ref{figModelRec}c depicts the maximal precipitation rate $r_+$ that can be predicted by our method for the different mismatch 
factors. More specifically, Fig.~\ref{figModelRec}c depicts the median (circles) and the 25\%-75\% quantile range (errorbars) based on 1000 realizations, as a function of the length of the simulated time series, $N$, in model's years (see also Figs. \ref{fig_predicted_precipitation_right_tail_gamma_sort} for the curves of the other 15 stations). $r_+$ appears to increase logarithmically as the length of the time series $N$ increases, a behavior that one may expect to observe due to the exponential tail of $P(r)$. The recurrence rate of $r_+$ as a function of the length of the time series is presented in Fig.~\ref{figModelRec}d and it appears to follow a power law with an exponent of $\sim-1$. Importantly, our method provides prediction beyond the range of the available data. The first point in Fig.~\ref{figModelRec}c,d presents the results for time series of lengths $N=72$ years (as in Fig.~\ref{figResultsModel}). The observed maximal precipitation for Sede-Boqer time series is less than 200 mm/yr---the predicted maximum annual precipitation is much larger with $r_+\simeq270$ mm/yr (for factor 2) corresponding to events with a recurrence time of about 1000 yrs, far beyond the 72 years of the time series. We elaborate more on the recurrence time and precipitation below.

\begin{figure*}[t]
\includegraphics[width=1\linewidth]{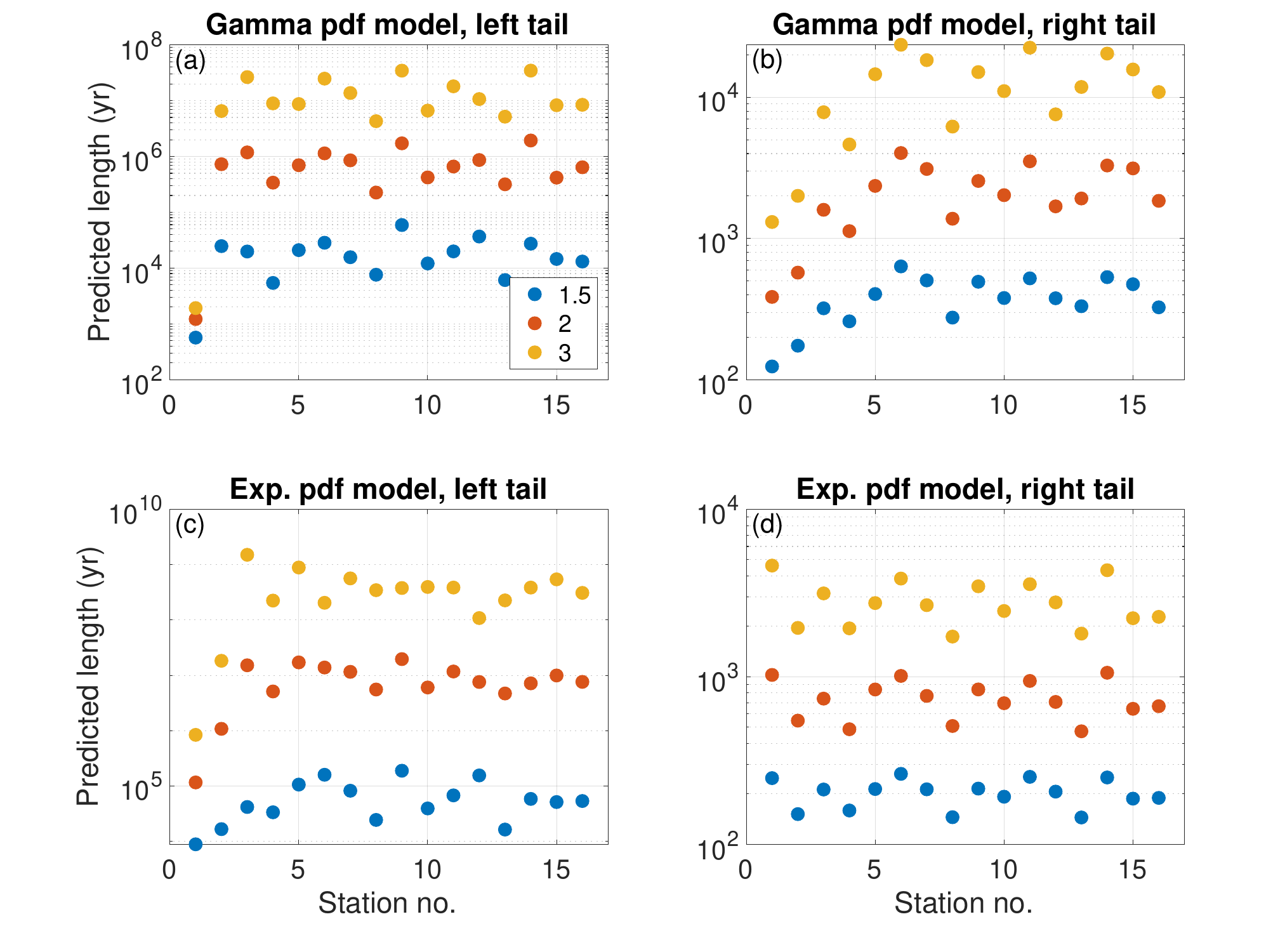}
\caption{The recurrence times (in years) of the annual precipitation rates $r_{\mp}$ at which the ratio between the theoretical and convoluted PDFs is equal to 1.5, 2, or 3 (blue, red, and yellow symbols); i.e., $P(r) / P_{\text{conv}}(r)$ or $P_{\text{conv}}(r) / P(r)$ equal to 1.5, 2, 3 (the crossing points between the solid and dashed lines in Fig.~\ref{figModelRec}b). The analysis was performed for the 16 stations under consideration based on the first point of the curves of Fig. \ref{figModelRec}d and Figs.~\ref{fig_recurrence_time_right_tail_gamma_sort}--\ref{fig_recurrence_time_left_tail_exp_sort}. (a) Left tail based on the gamma PDF model, (b) right tail based on the gamma PDF model, (c) left tail based on the exponential PDF model, (d) right tail based on the exponential PDF model. The stations are ordered according to the mean annual precipitation and it is apparent that: (i) the predicted length increases with the precipitation rate in panels b and d, (ii) as expected larger ratios yield longer prediction time, and (iii) the dry years' prediction time (left tail) is much longer than the wet years' prediction time (right tail). Station no. two (Sde Boker) is the station analyzed in the main text. 
}
\label{fig_predicted_time}
\end{figure*}

We found that, for all other 15 stations under consideration, the recurrence rate of $r_+$ (in 1/yr) versus the length $N$ of the simulated time series (in years) approximately follows a power law too (Fig.~\ref{fig_recurrence_time_right_tail_gamma_sort},\ref{fig_recurrence_time_right_tail_exp_sort}). 
We estimated the power law exponents for all the 16 stations under consideration and found that different stations are characterized by different exponents, within the range $[-2.5,-1]$ (Fig.~\ref{figExponents}a). The stations are sorted according to the annual mean precipitation and it is apparent that the exponent is more negative as the precipitation rate increases. In addition, the larger mismatch factors (represented by different colors) result in a more negative exponent. We also studied an additional model in which the monthly rainfall is modeled by an exponential distribution (instead of the gamma distribution). The sum of exponential PDF is the hypoexponential distribution; we performed the same analysis shown in Fig.~\ref{figModelRec} using the hypoexponential distribution as the theoretical distribution and found the corresponding power law exponents (Fig.~\ref{figExponents}b). Here the exponents span a smaller range of $[-1.6,-1]$ where we did not observe a clear relation between the annual mean precipitation and the exponent. Yet, also here larger mismatch factors result in a more negative exponent. The individual curves for each of the stations from which we calculated the exponents presented in Fig.~\ref{figExponents} are shown in Figs.~\ref{fig_predicted_precipitation_right_tail_gamma_sort},\ref{fig_predicted_precipitation_right_tail_exp_sort}. We note that the model for which the monthly rainfall is gamma-distributed exhibits better performance than the model for which the monthly rainfall is exponentially distributed.  
%
The recurrence time of $r_+$ for the gamma and exponential PDF models using ratios of 1.5, 2, and 3, for the 16 stations under consideration is depicted in Fig.~\ref{fig_predicted_time}a,c. 

The left tail of the distribution exhibits very different behavior in comparison to the right tail; see Fig.~\ref{figModelRec}a,b and Figs. \ref{fig_recurrence_time_left_tail_gamma_sort}, \ref{fig_recurrence_time_left_tail_exp_sort} in comparison to Figs. \ref{fig_recurrence_time_right_tail_gamma_sort}, \ref{fig_recurrence_time_right_tail_exp_sort}. This is expressed in the predictions in the left tail which are remarkably accurate, even for very rare events; see Fig.~\ref{fig_predicted_time}a,c versus Fig.~\ref{fig_predicted_time}b,d. 
In the left tail, our method provides accurate predictions up to a minimal annual precipitation rate $r_-$ whose recurrence time
exceeds several thousand years, much longer than the length of the simulated time series 
\footnote{Note that the lengths of the simulated time series are the same as the lengths of the observed time series.}
of $\sim100$ years. This is valid for both gamma and exponential PDF models, and even when $r_-$ is calculated using a relatively small mismatch factor of 1.5. Notably, except for station no.~1 (Elat) which is characterized by a very low annual precipitation rate of less than 30 mm/yr and for which the precipitation at some years was only a few mm (Fig.~\ref{fig_stations_time_series}), the prediction time of $r_-$ for the other stations exceeds 5000 years. The prediction time for a rare dry year for station no.~1 is relatively not long simply because such a year is more probable in this station than for the other stations.
 Furthermore, Fig.~\ref{figModelRec}d indicates that the ratio for the left tail is less than one. This is valid for the extreme arid station of Sde-Boqer while for other, less arid, stations this ratio is not necessarily smaller than one. 

We note that the prediction time for a rare wet year is longer for the station with higher annual mean precipitation (Fig.~\ref{fig_predicted_time}b).
From Fig.~\ref{fig_predicted_time} we find that, for a mismatch of a factor of two, the predicted recurrence times, both for wet and dry years ($r_{\pm}$), are much longer than the length of the observed annual precipitation time series; i.e., the length of the time series of the 16 stations we consider is around 100 years while the predicted recurrence times are at least 400 years, for both models and both tails. As expected, the smaller ratio of 1.5 yielded a shorter prediction time, yet, larger than the length of the observed time series. For a mismatch factor 3 (yellow circles) the prediction time is much longer than the length of the original time series, more than 1000 years. 
%
In Fig.~\ref{fig_predicted_precipitation} we plotted the minimum/maximum predicted annual precipitations $r_{\mp}$ versus the corresponding minimum/maximum annual precipitations observed in the simulated data. For both gamma and exponential PDF models, and for mismatch ratios of 1.5, 2, and 3, the maximum predicted annual precipitation is larger than the maximum observed precipitation for almost all stations. The situation is clearer for the minimum precipitation rate where the predicted minimum annual precipitation is well below the observed value, consistent with the much longer prediction time for minimum precipitation. An exception is station no. 1 (Elat) for which a minimum zero annual precipitation is a probable scenario as this station experienced a few mm annual precipitation for some years (Fig. \ref{fig_stations_time_series}). 

We thus conclude that the method we proposed not only provides prediction for the central part of the annual rain distribution but also provides reliable predictions far outside the range of the actual data, especially in the left tail of the distribution. 
As explained above, the difference in the quality of our predictions in the two tails follows from the qualitative difference between their behaviors. The right tail of the annual rain distribution (wet years) is not bounded such that a longer time series yields a longer range of ratio that is close to one (Fig.~\ref{figModelRec}b). The left tail is bounded by $r=0$ and the annual precipitation time series of very different lengths have almost identical left tails and mismatch ratios (Fig.~\ref{figModelRec}a,b and Figs.~\ref{fig_recurrence_time_left_tail_gamma_sort}, \ref{fig_recurrence_time_right_tail_gamma_sort}).

\section{Discussion and Conclusion}

Precipitation plays an important role in the climate system. As such, it is necessary to develop reliable statistical methods to analyze the precipitation data. These can be used to quantify the statistical properties of the data, as well as, to provide better predictions for future events. This is especially important in light of the ongoing climate change which is expected to experience more frequent rare events (droughts and floods) \cite{Masson-Delmotte-Zhai-Pirani-et-al-2021:climate,Pörtner-Roberts-Adams-et-al-2022:climate}. We aimed to improve the prediction of the distribution of annual rain. In particular, we aimed to provide reliable statistical predictions of rare dry or wet years, including those that fall outside the range of the observed annual rain.

The proposed method is based on the approximation that the rainfall of different months is independent---this assumption is supported by the data. We then regarded the rainfall of the different months as independent random variables and constructed the annual rainfall distribution by convoluting the measured monthly PDFs. This is in contrast to previous studies that assumed that the annual rainfall follows a specific distribution (typically the gamma distribution) and fitted the distribution to the annual precipitation data. The method we proposed provides reliable estimation for the central part of the distribution as well as outside the range of the measured data. 
In particular, the method we propose predicts the likelihood of extremely dry years remarkably accurately, including events whose recurrence time is very long. Our method also predicts the likelihood of unusually (but not extremely) wet years. In contrast to our analysis, previous studies predicted extremely wet or dry years based on the tails of assumed distributions--as the tails of the real distribution are not necessarily related to the central part of the distribution, such predictions may be unreliable with a mismatch of orders of magnitudes.

While our predictions yield very large recurrence times $T$ (e.g., of the order of tens of thousands of years) for certain stations/cases, they should of course not be interpreted as an estimate of the typical time that will elapse until a wet or dry year will happen or already happened. This is because it is known that the climate system exhibits a very long temporal correlations of tens of thousands of years (like the glacial-interglacial oscillations), and, in fact, these correlations are stronger for longer times \cite{Pelletier-1997:analysis}. Instead, a recurrence time $T$ should be interpreted as a probability of $1/T$ (where $T$ is measured in years) that the wet or dry year in question will occur in a given year with today's climate.

The method we proposed here can be applied to predict probabilities of events involving climate variables other than rainfall, e.g., temperature (including prolonged heatwaves or cold spells) and winds. Moreover, one could apply it to other dynamical systems, when considering statistical fluctuations of quantities that are integrated or averaged over a period that is much longer than the correlation time of the dynamical system in question \cite{Touchette-2009:large}.

 In the present study we focused on precipitation in Israel, covering a wide range of precipitation rate, from extreme arid to semi arid environment. Precipitation in Israel falls during the winter. The conclusions of the present study are thus valid for the Israeli environment (or similar environments) but may be different for other environments for which the precipitation of different months may be correlated. In arid regions, like Israel, the prediction of extremely dry years is probably more important than the prediction of extremely wet years. However, the prediction of high precipitation rate is probably more important in regions with high precipitation rate, which are prone to flooding in populated areas. Applying the proposed method to temperature time series may not be straightforward due to temporal correlations. Yet, there are several several daily temperature records that extend more than two centuries that may strengthen the statistical significance of the results.

A basic limitation that our method has is its assumption that the monthly rainfalls are uncorrelated. Although this is a good approximation, it would be interesting to improve our method by taking the existing possible small correlations into account, and/or by changing the size of the time windows. The climate system exhibits long-term fluctuations or trends (such as the global warming trend) and the extension of our method to handle these conditions may be important to assess the impact of phenomena like global warming on the distribution and extreme events of different climate variables. 

\subsection*{Acknowledgments}
We acknowledge helpful discussions with Dorian Abbot, Jonathan  Wear, Valerio Lucarini, Michael Wilkinson, and Amos Zemel. NRS acknowledges support from the Israel Science Foundation (ISF) through Grant No. 2651/23. 


%

\setcounter{figure}{0}
\renewcommand{\thefigure}{S\arabic{figure}}

\setcounter{table}{0}
\renewcommand{\thetable}{S\arabic{table}}

\appendix
\section{Supplemental table and figures}

In this appendix, we present some additional details that support the findings reported in the main text.

\begin{table*}
    \centering
    \begin{tabularx}{\textwidth}{c|c|l|c|>{\centering\arraybackslash}X|>{\centering\arraybackslash}X|>{\centering\arraybackslash}X|>{\centering\arraybackslash}X|>{\centering\arraybackslash}X}
    \hline
No. & IMS st. no. & St. name & Long. ($^\circ$E) & Lat. ($^\circ$N) & First year & Last year & No. of years & Mean annual rain (mm/yr) \\
\hline
1 & 347700 & Elat & 34.9542 & 29.5526 & 1950 & 2023 & 74 & 27.9 \\
2 & 253000 & Sede Boqer & 34.795 & 30.8702 & 1952 & 2023 & 72 & 90.5 \\
3 & 321850 & Tirat Zevi & 35.5258 & 32.4222 & 1939 & 2023 & 84 & 274.1 \\
4 & 247550 & Bet Guvrin & 34.8933 & 31.6136 & 1951 & 2023 & 73 & 402.4 \\
5 & 221450 & Merhavya & 35.3075 & 32.6028 & 1921 & 2017 & 88 & 453.4 \\
6 & 246550 & Beit Jimal & 34.9762 & 31.7248 & 1920 & 2023 & 104 & 498.1 \\
7 & 220850 & Mizra & 35.2862 & 32.6523 & 1929 & 2023 & 91 & 503.1 \\
8 & 120900 & Atlit & 34.9389 & 32.7057 & 1929 & 2023 & 73 & 515.2 \\
9 & 242950 & Latrun Monastry & 34.98 & 31.8324 & 1901 & 2023 & 107 & 517.2 \\
10 & 220550 & Ginnegar & 35.2563 & 32.6624 & 1927 & 2023 & 84 & 524 \\
11 & 244850 & Jerusalem St Anne & 35.2363 & 31.7808 & 1909 & 2023 & 97 & 525.8 \\
12 & 136650 & Miqwe Yisrael & 34.7846 & 32.0318 & 1908 & 2023 & 102 & 538.9 \\
13 & 110050 & Kefar Rosh Haniqra & 35.1149 & 33.0861 & 1950 & 2023 & 71 & 601.6 \\
14 & 243500 & Qiryat Anavim & 35.1199 & 31.8098 & 1923 & 2023 & 95 & 671.3 \\
15 & 211900 & Zefat Har Kenaan & 35.507 & 32.98 & 1940 & 2023 & 81 & 708.8 \\
16 & 120750 & Yagur & 35.0783 & 32.7403 & 1929 & 2023 & 85 & 710.9 \\    
\hline
\end{tabularx}
    \caption{The details of the 16 stations analyzed in this study. The rainfall data was downloaded from the Israeli Meteorological Survey (IMS) web page (\href{https://ims.gov.il/en/data_gov}{https://ims.gov.il/en/data\_gov} or \href{https://data.gov.il/dataset/481}{https://data.gov.il/dataset/481}). Note that for some stations the data for some years is missing. Annual rains in Israel accumulate from the beginning of September to the end of August; e.g., the rain of 2023 is the sum of precipitation that occurred between Sep. 1, 2022, and August 31, 2023. The results that are shown in the main text are those of station no. 2 (Sede Boqer).}
    \label{tab:station_details}
\end{table*}

\begin{figure*}[t]
\includegraphics[width=1\linewidth]{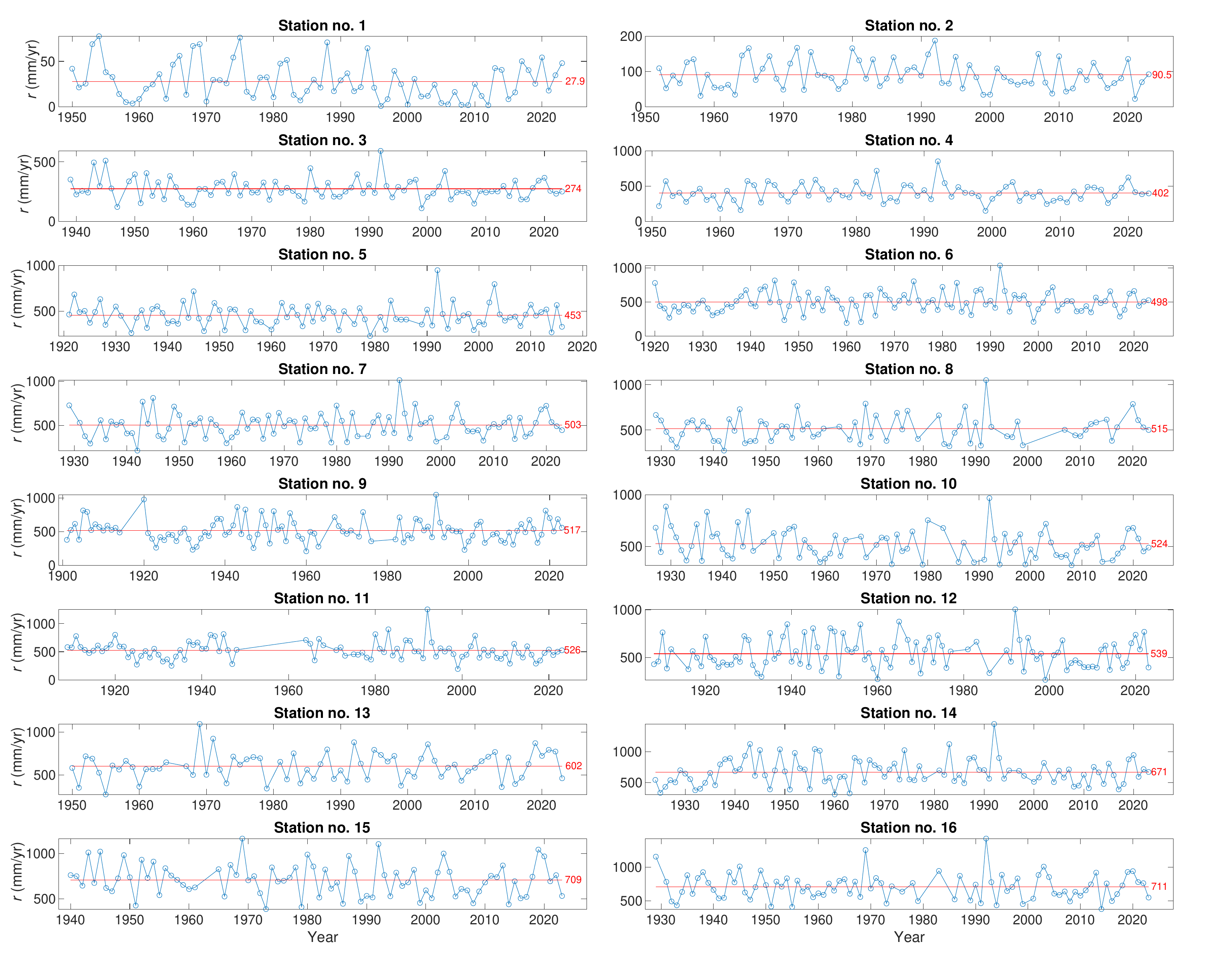}
\caption{Annual precipitation rate versus time (year) for the 16 stations analyzed in this study (see Table \ref{tab:station_details}). The horizontal red line indicates the mean precipitation rate (in mm/yr). The main text presents the results of station no. 2.
}
\label{fig_stations_time_series}
\end{figure*}

\begin{figure*}[t]
\includegraphics[width=1\linewidth]{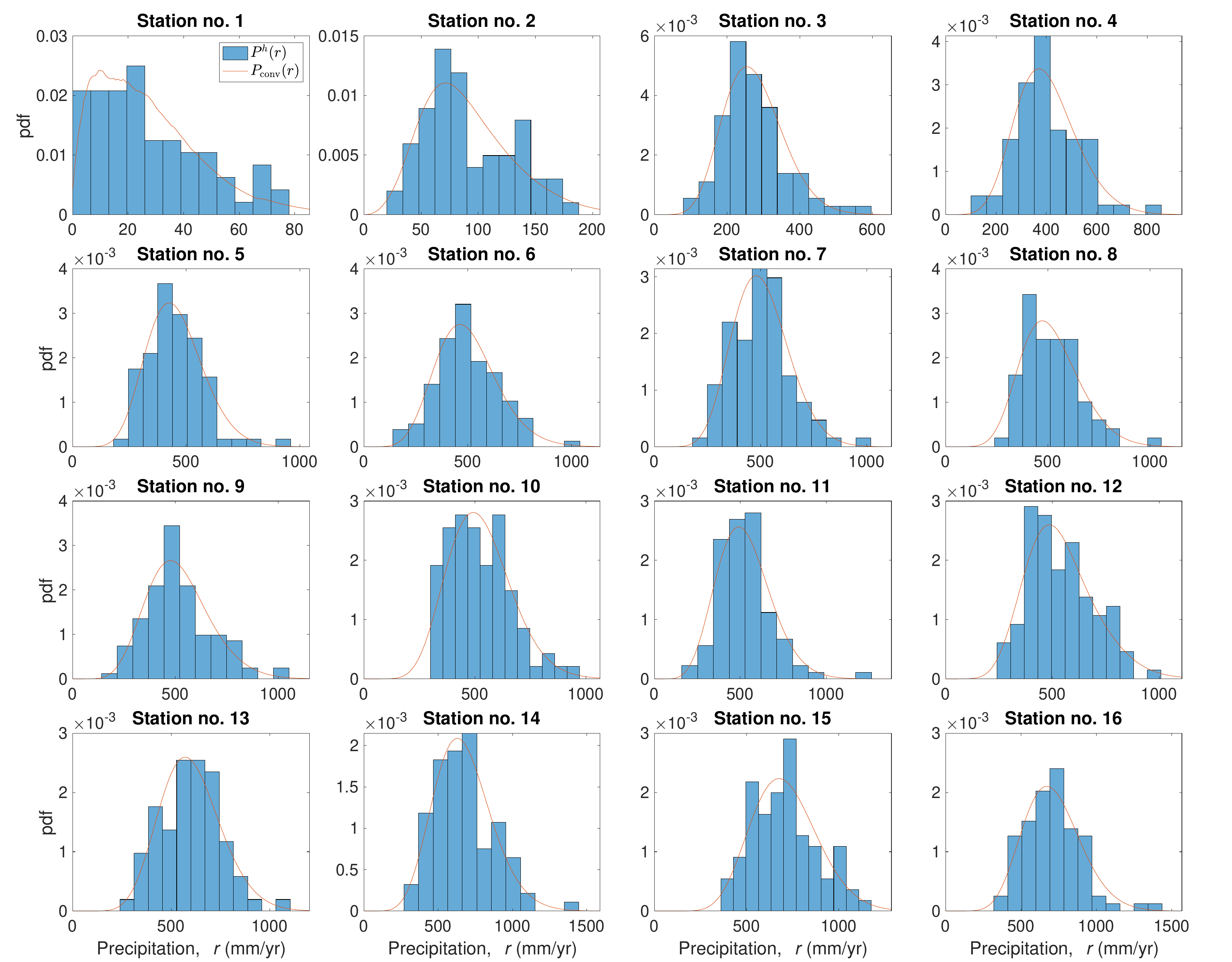}
\caption{The histogram (blue bars), $P^h(r)$, and convolution PDF (red curve), $P_{\rm conv}(r)$, of the 16 annual precipitation time series presented in Fig.~\ref{fig_stations_time_series}.
}
\label{fig_stations_pdf}
\end{figure*}

\begin{figure*}
\centering\includegraphics[width=0.9\linewidth]{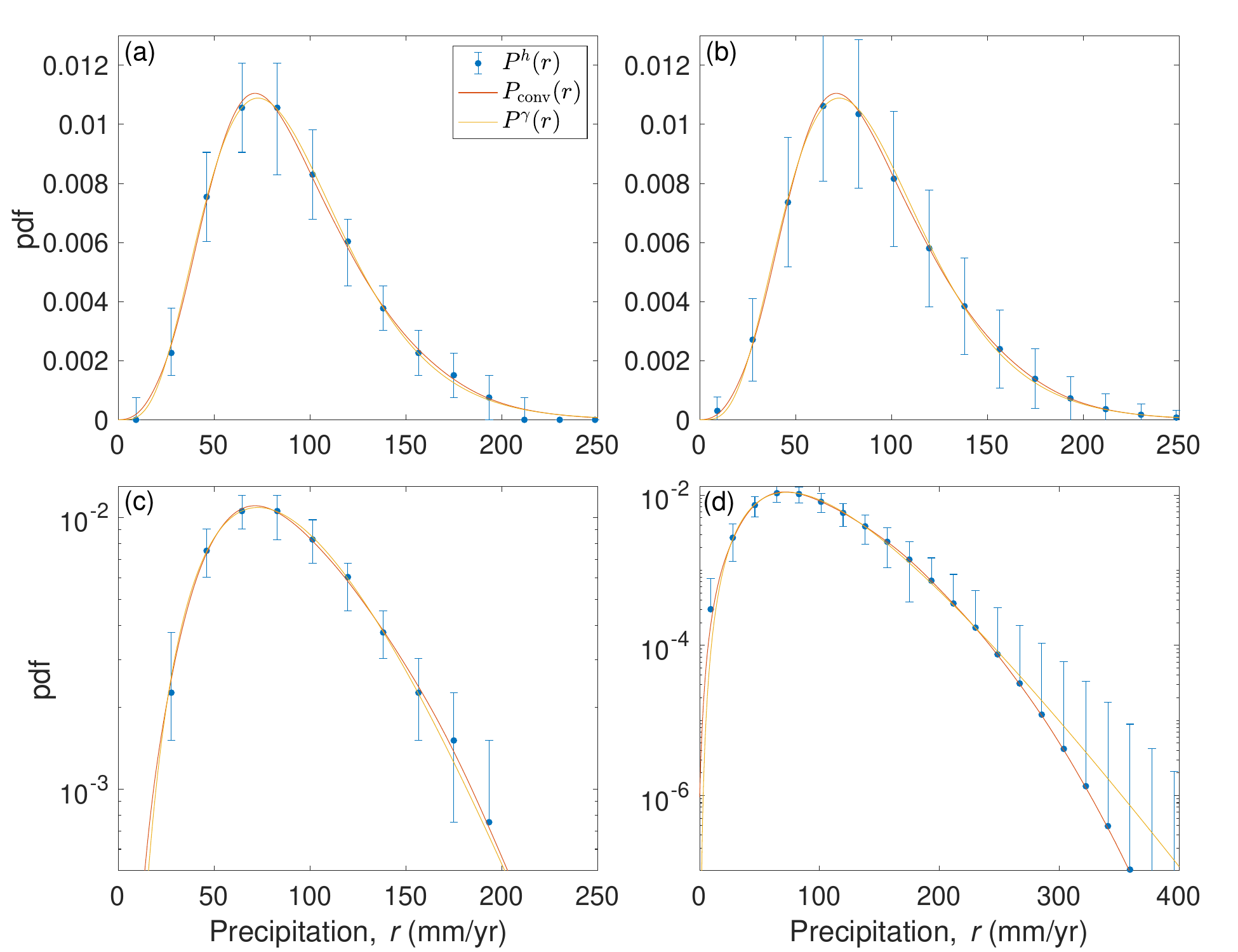}
\caption{The results of bootstrapping of Sde-Boqer station based on 13888888 sets of 72 years (all together slightly less than one billion years). (a) The median $P^h(r)$ (blue circles) and the 25\%-75\% quantiles (error bars) of the bootstrapping sets, together with the convolution pdf, $P_{\rm conv}(r)$ (red) and the gamma pdf, $P^\gamma(r)$ (yellow). (b) Same as a depicting the mean and the standard deviation. (c) Same as a in semi-log plot. (d) Same as b in semi-log plot. Note the large error bars. Also note that in panels c, d some of the symbols or error bars are now shown as they are equal or smaller than zero.}  
\label{fig:bootstrapping}
\end{figure*}

\begin{figure*}[t]
\includegraphics[width=1\linewidth]{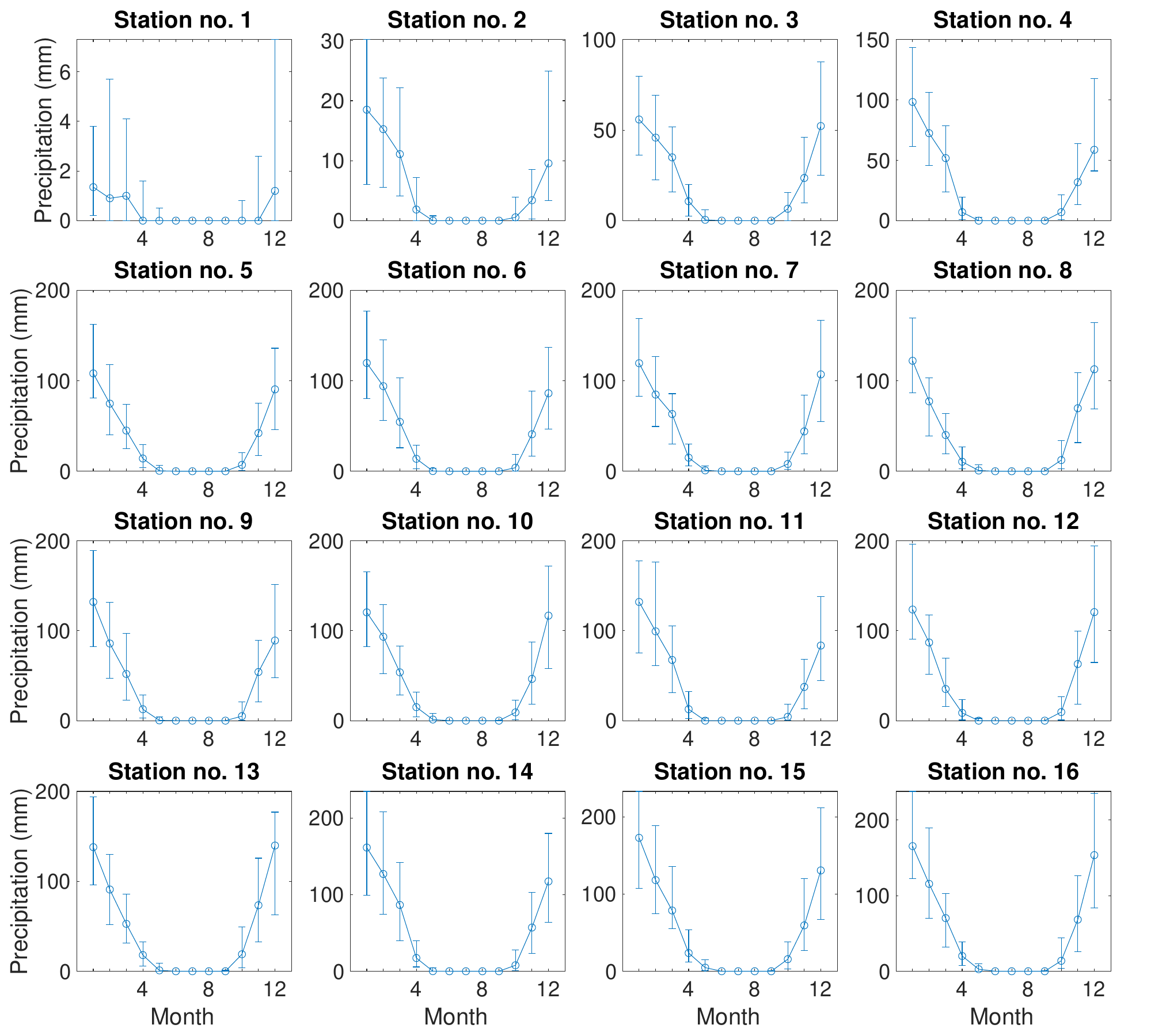}
\caption{
Monthly median (circles) and 25\%-75\% quantile range (errorbars) for the 16 stations under consideration (see Table \ref{tab:station_details}). Note the dry summer months.
}
\label{figMontlhyRain}
\end{figure*}

\begin{figure*}[t]
\includegraphics[width=0.9\linewidth]{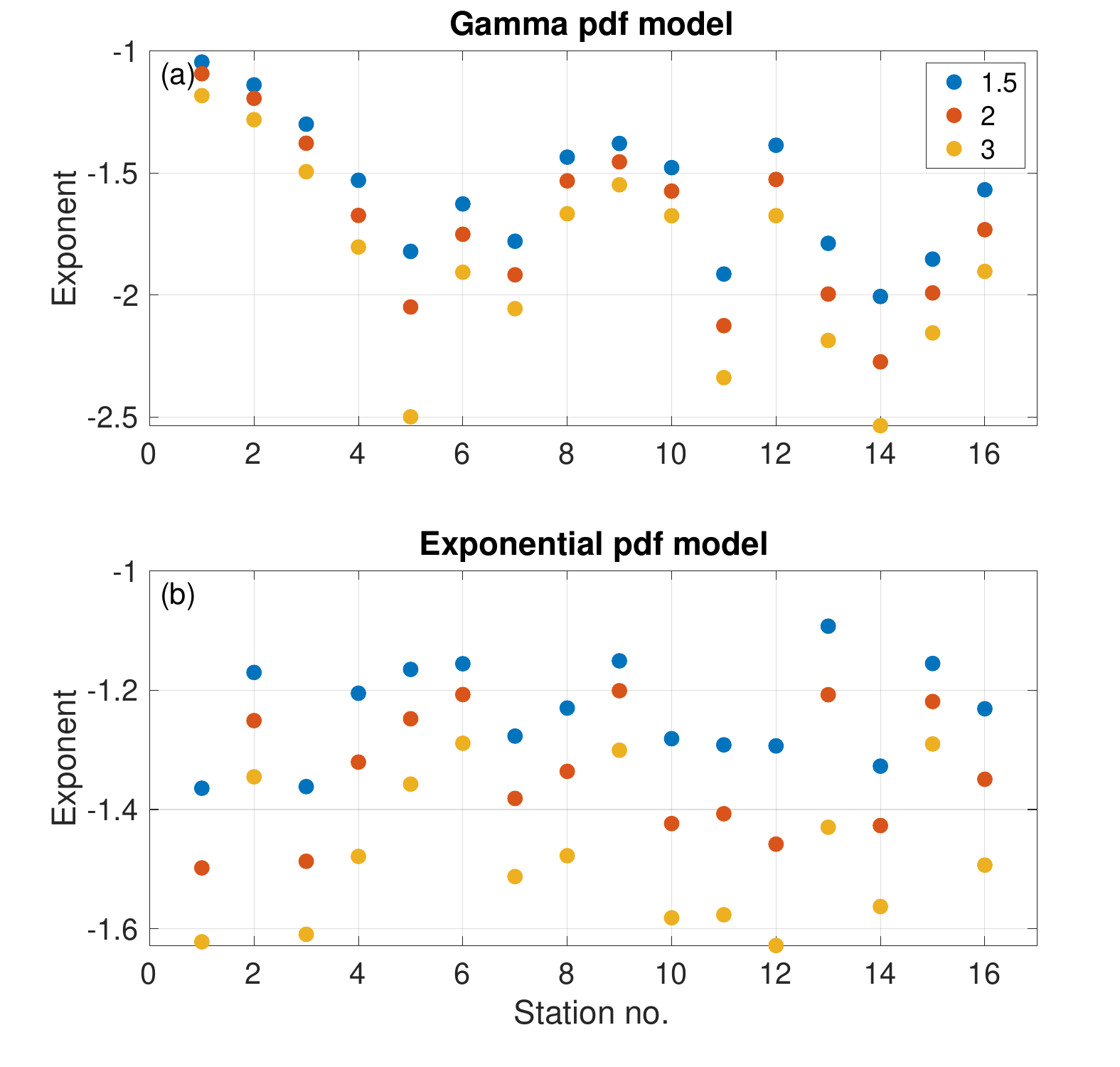}
\caption{
The slope (on a log-log plot) of the recurrence rate curve presented in Fig. \ref{figModelRec}d and Figs.~\ref{fig_predicted_precipitation_right_tail_gamma_sort}, \ref{fig_predicted_precipitation_right_tail_exp_sort}, for the 16 stations analyzed in this study; the different colors indicate different mismatch factors (1.5, 2, and 3). We used two different models for the annual precipitation rate: the monthly rain is assumed to follow the (a) gamma distribution and the (b) exponential distribution. The distribution parameters were calculated using MLE of the monthly rain of the different stations. Note that: (i) different stations are characterized by different exponents, (ii) the model that is based on the exponential distribution of monthly rain results in a larger (less negative) exponent for most of the stations, (iii) in panel a, stations with larger mean precipitation rate are characterized by more negative exponent, and (iv) larger mismatch factors result in larger (more negative) exponents. 
}
\label{figExponents}
\end{figure*}

\begin{figure*}[t]
\includegraphics[width=1\linewidth]{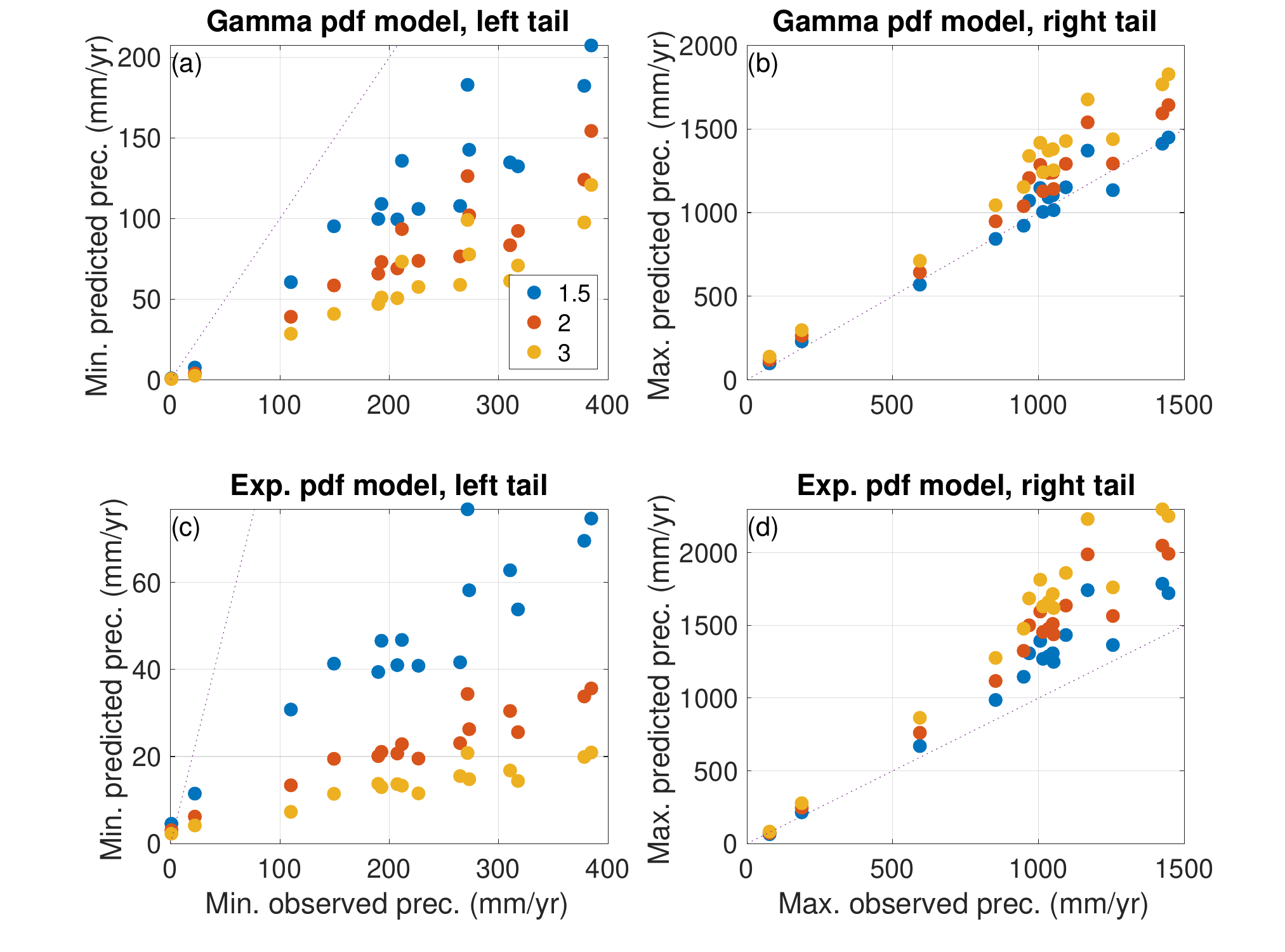}
\caption{The (minimum and maximum) predicted annual precipitation rates $r_{\mp}$ (in mm/yr) at which the mismatch factor between the theoretical and convoluted PDFs is equal to 1.5, 2, or 3 (blue, red, and yellow symbols) as a function of the minimum and maximum observed annual precipitation; i.e., $P(r) / P_{\text{conv}}(r)$ or $P_{\text{conv}}(r) / P(r)$ equal to 1.5,2,3 (the crossing points between the solid and dashed lines in Fig.~\ref{figModelRec}b). The analysis was performed for the 16 stations under consideration based on data like that of the first point of the curves of Fig.~\ref{figModelRec}c; see Figs.~\ref{fig_predicted_precipitation_right_tail_gamma_sort}--\ref{fig_predicted_precipitation_left_tail_exp_sort}. (a) Left tail (minimum annual precipitation) based on the gamma PDF model, (b) right tail (maximum annual precipitation) based on the gamma PDF model, (c) left tail based on the exponential PDF model, (d) right tail based on the exponential PDF model. The black dotted line in each panel indicates the line at which the predicted (minimum or maximum) annual precipitation should be equal to the observed (minimum or maximum) annual precipitation. Almost all points lay below (above) the black dotted line in panels a and c (b and d) indicating that the predicted annual precipitation is smaller (larger) than the observed minimum (maximum) precipitation. In addition, as expected, larger ratios yield a more pronounced difference between the observed and the predicted minimum/maximum precipitation. Station no. two (Sde Boker) is the station analyzed in the main text.
}
\label{fig_predicted_precipitation}
\end{figure*}

\begin{figure*}[t]
\includegraphics[width=1\linewidth]{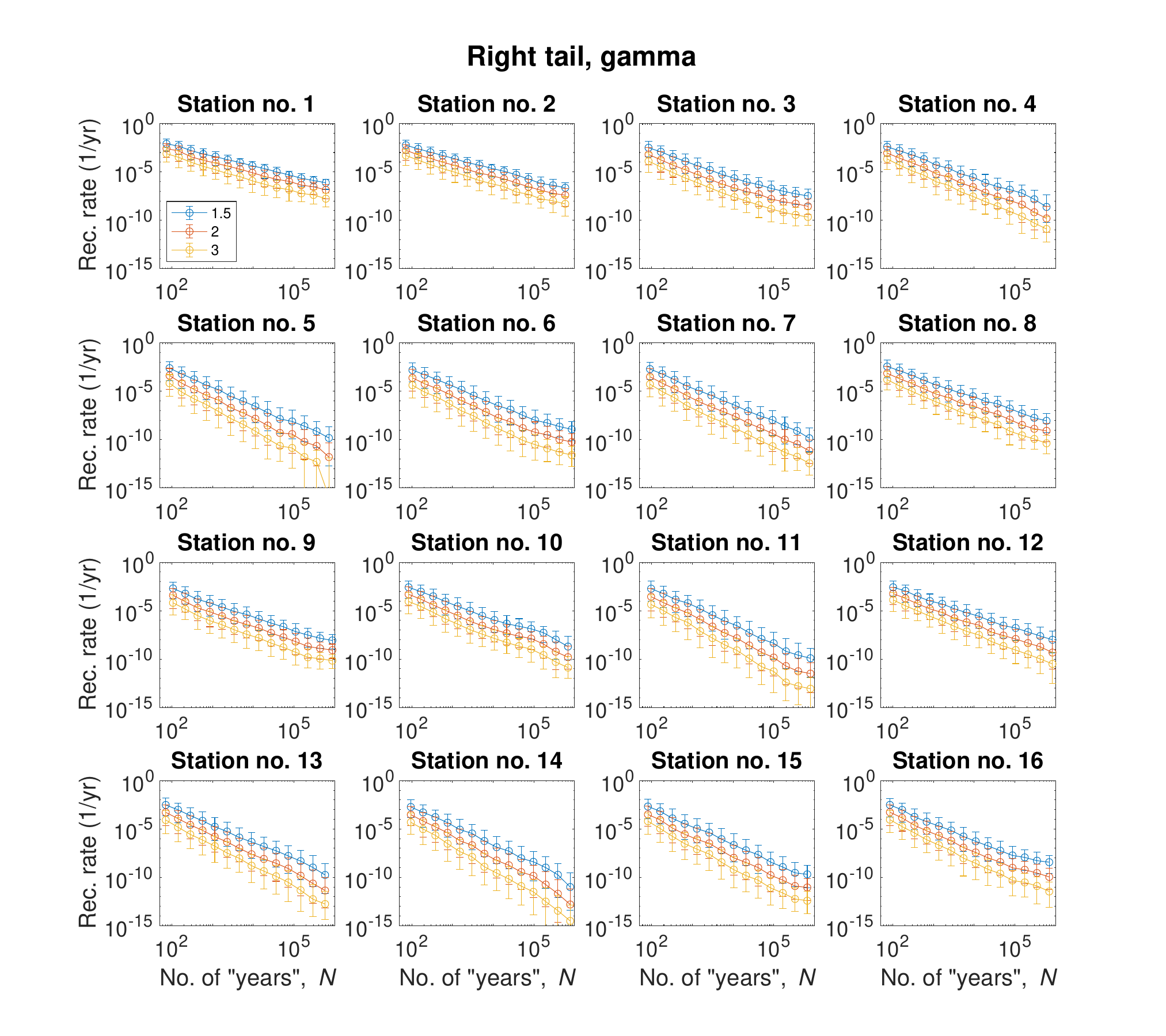}
\caption{The recurrence rate (in 1/yr, see Fig.~\ref{figModelRec}d) of the maximal predicted precipitation $r_+$ versus the length of the model's time series (in years) for 1.5 (blue), 2 (red), and 3 (yellow) mismatch factors on a log-log plot for the 16 stations considered in this study (see Table~\ref{tab:station_details}). The symbols indicate the median and the errorbars indicate the 25\%-75\% quantiles of 1000 model realizations. The figure depicts the results of the gamma PDF model based on the right tail of the distribution; i.e., providing prediction for the recurrence rate for maximum annual rain. 
}
\label{fig_recurrence_time_right_tail_gamma_sort}
\end{figure*}

\begin{figure*}[t]
\includegraphics[width=1\linewidth]{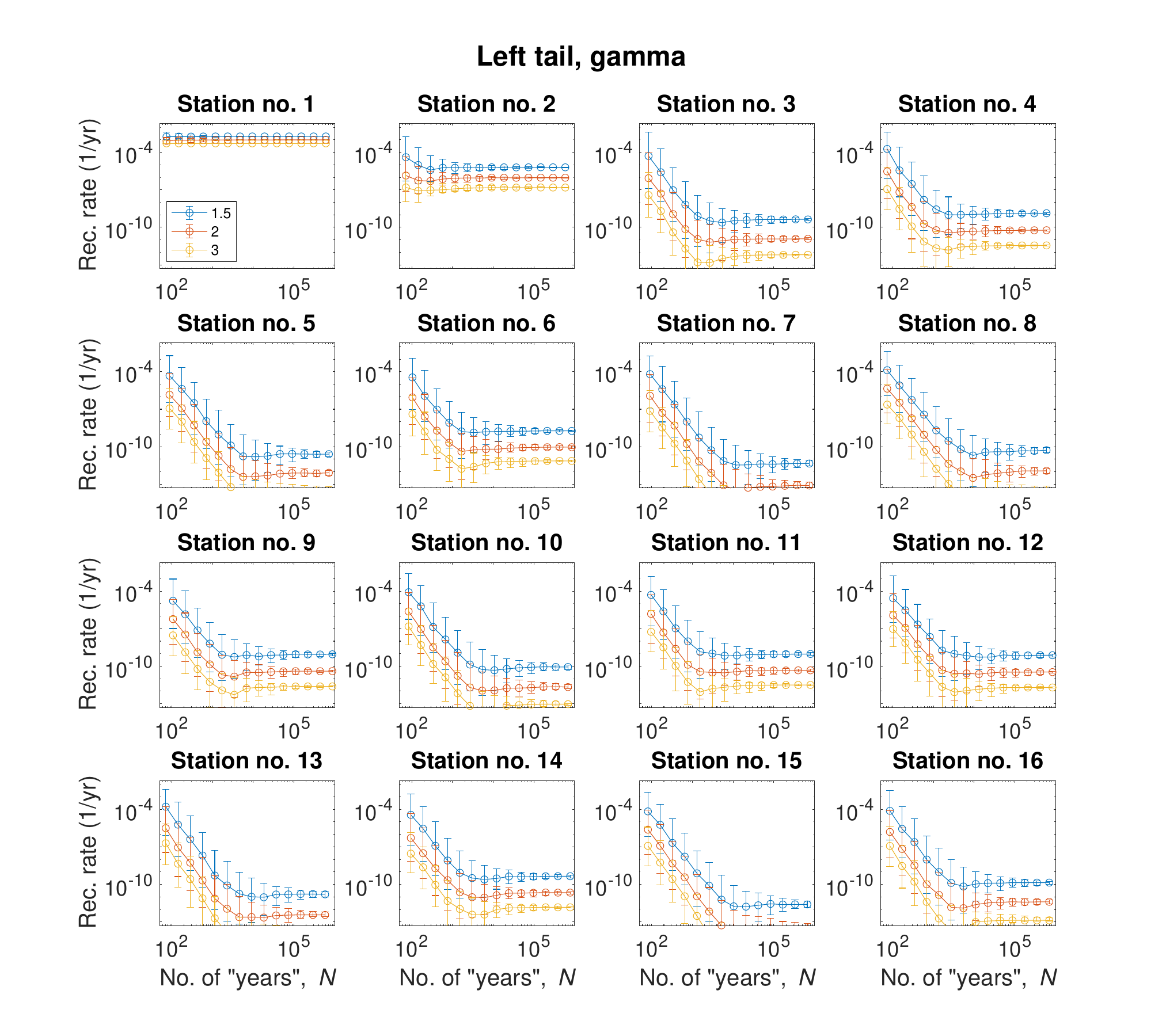}
\caption{Same as Fig.~\ref{fig_recurrence_time_right_tail_gamma_sort} for the left tail of the distribution; i.e., providing prediction for the recurrence rate for minimum annual rain $r_-$.
}
\label{fig_recurrence_time_left_tail_gamma_sort}
\end{figure*}

\begin{figure*}[t]
\includegraphics[width=1\linewidth]{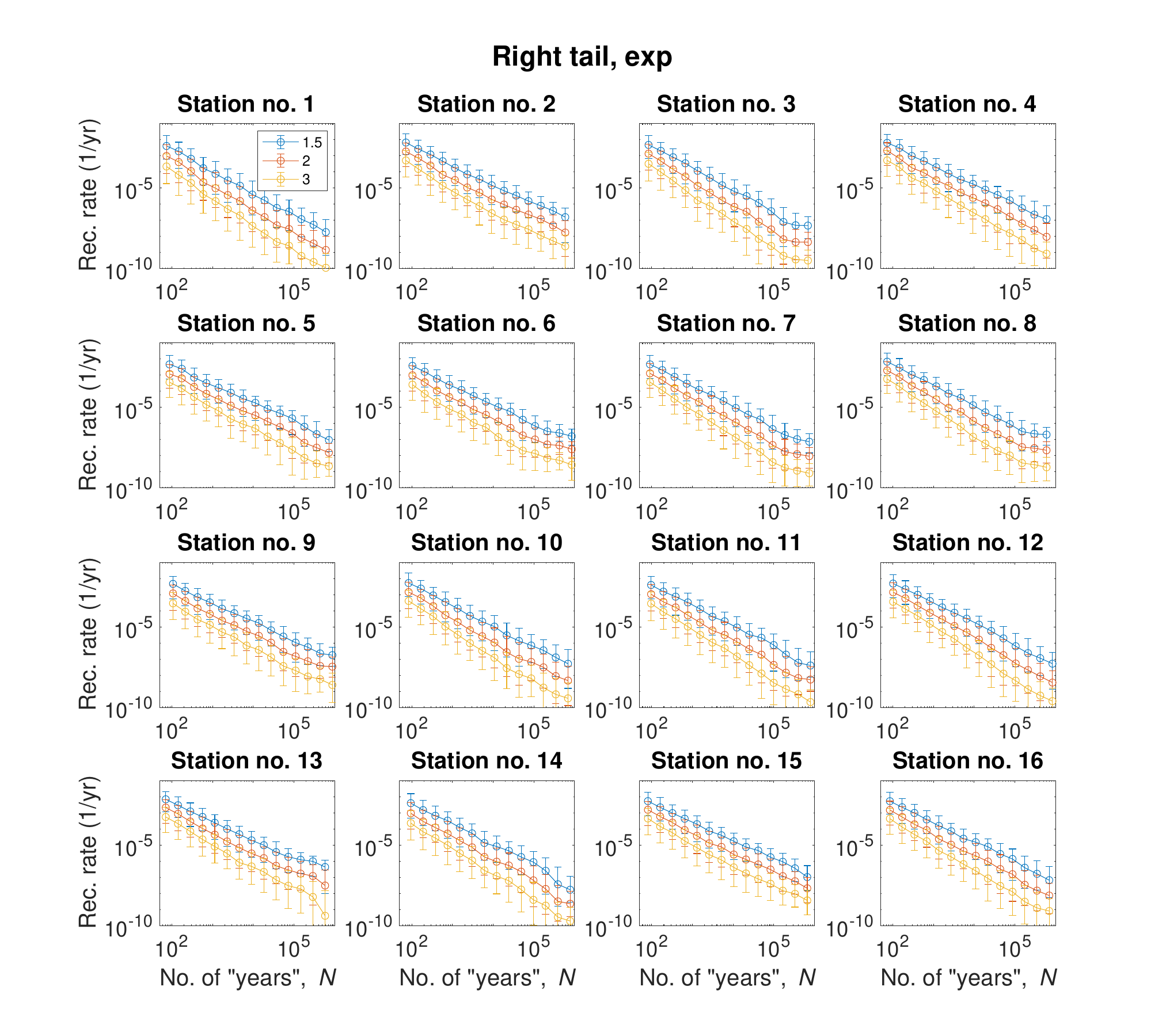}
\caption{Same as Fig.~\ref{fig_recurrence_time_right_tail_gamma_sort} but for the exponential PDF model.
}
\label{fig_recurrence_time_right_tail_exp_sort}
\end{figure*}

\begin{figure*}[t]
\includegraphics[width=1\linewidth]{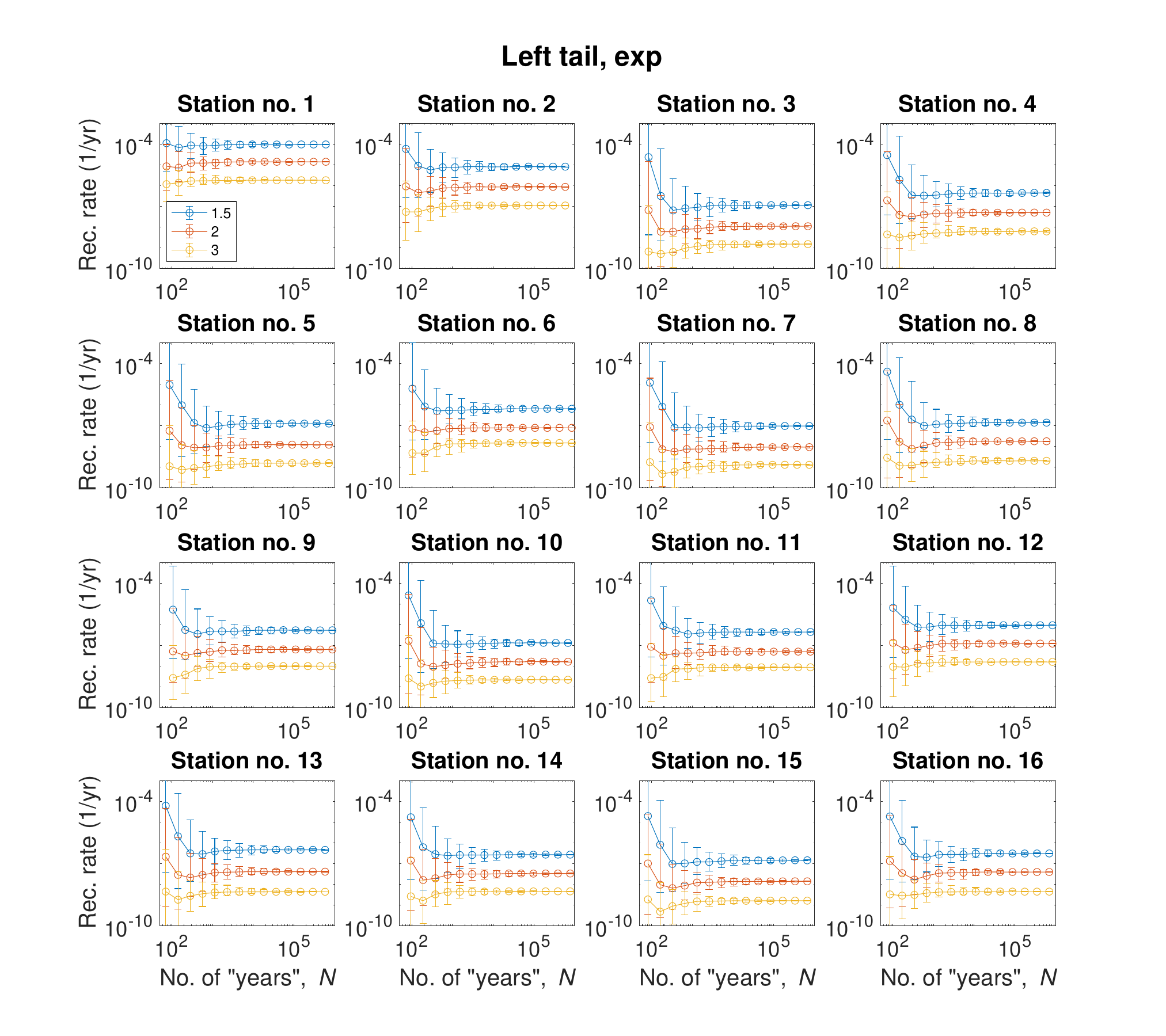}
\caption{Same as Fig.~\ref{fig_recurrence_time_right_tail_exp_sort} for the left tail of the distribution. 
}
\label{fig_recurrence_time_left_tail_exp_sort}
\end{figure*}

\begin{figure*}[t]
\includegraphics[width=1\linewidth]{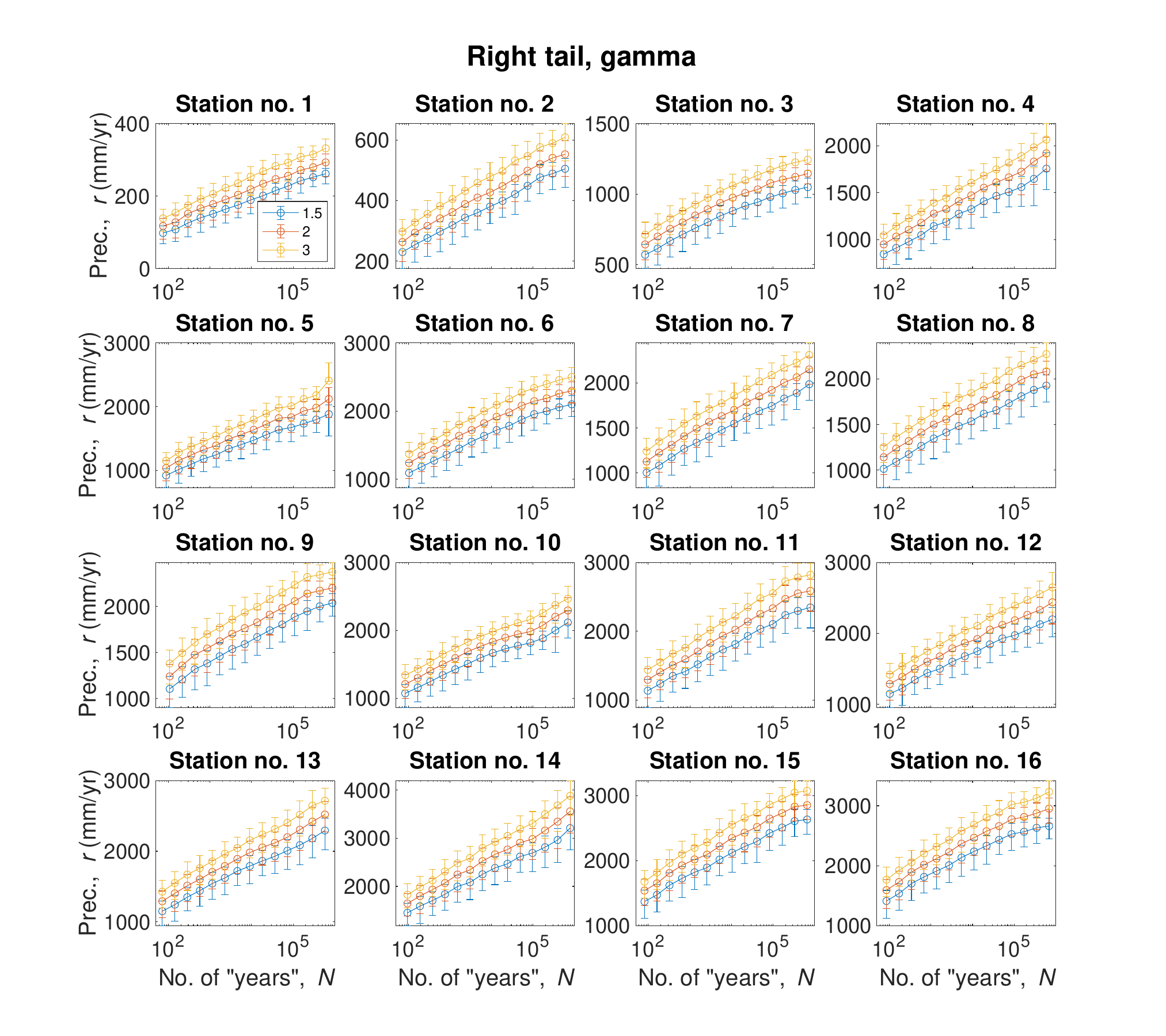}
\caption{The annual precipitation at which the mismatch factor exceeds 1.5 (blue), 2 (red), and three (yellow) versus the length of the time series (in model's years). The results are based on 1000 model realizations where the errorbars indicate the 25\%-75\% quantile range and the circles indicate the median; see Fig.~\ref{figModelRec}c. The figure depicts the results of the 16 stations considered in this study (see Table~\ref{tab:station_details}). The results presented here are based on the gamma PDF model, on the right tail of the distribution; i.e., the maximal annual predicted precipitation $r_+$.  
}
\label{fig_predicted_precipitation_right_tail_gamma_sort}
\end{figure*}

\begin{figure*}[t]
\includegraphics[width=1\linewidth]{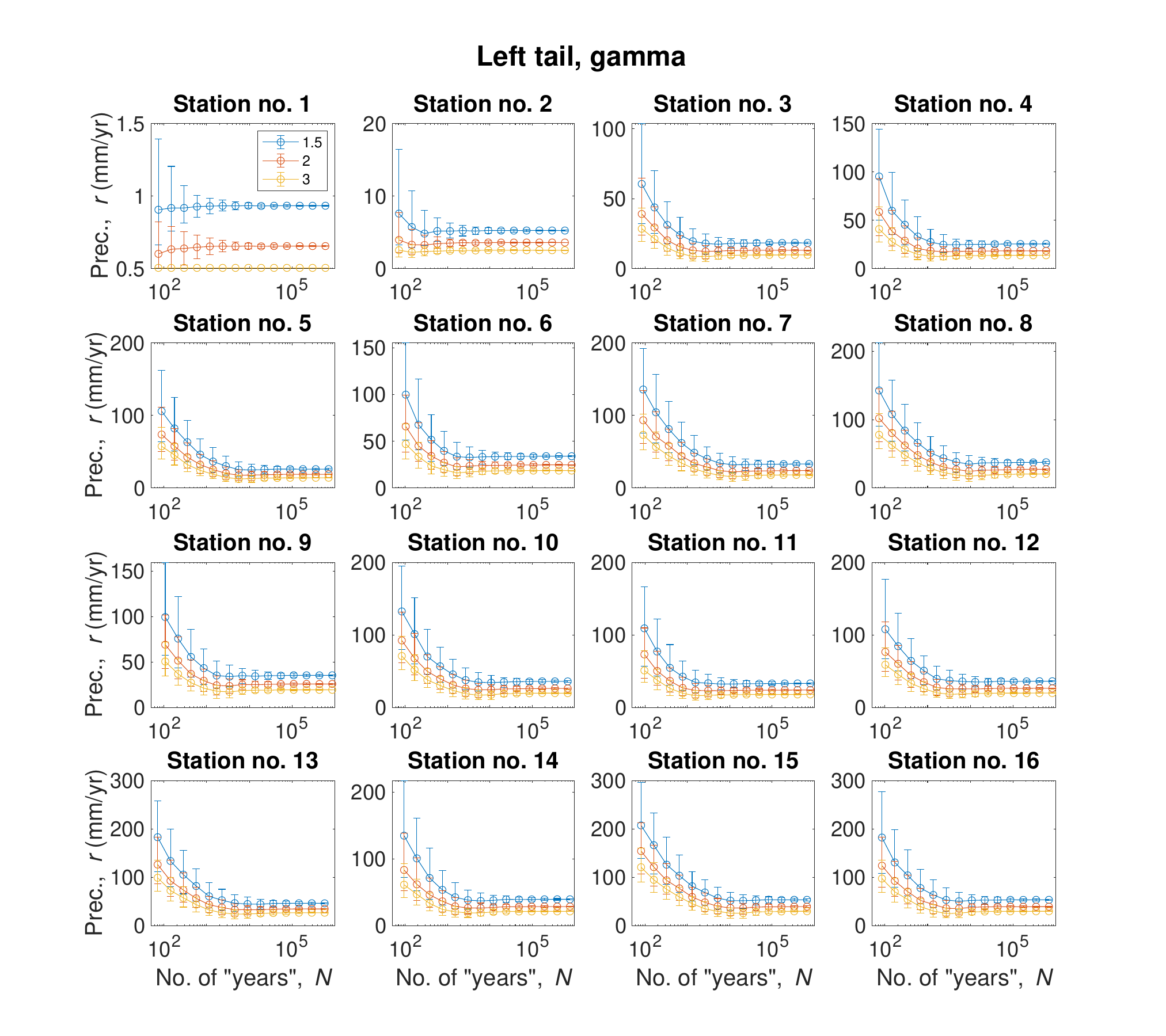}
\caption{Same as Fig.~\ref{fig_predicted_precipitation_right_tail_gamma_sort} for the left tail of the distribution; i.e., minimal predicted annual precipitation $r_-$.
}
\label{fig_predicted_precipitation_left_tail_gamma_sort}
\end{figure*}

\begin{figure*}[t]
\includegraphics[width=1\linewidth]{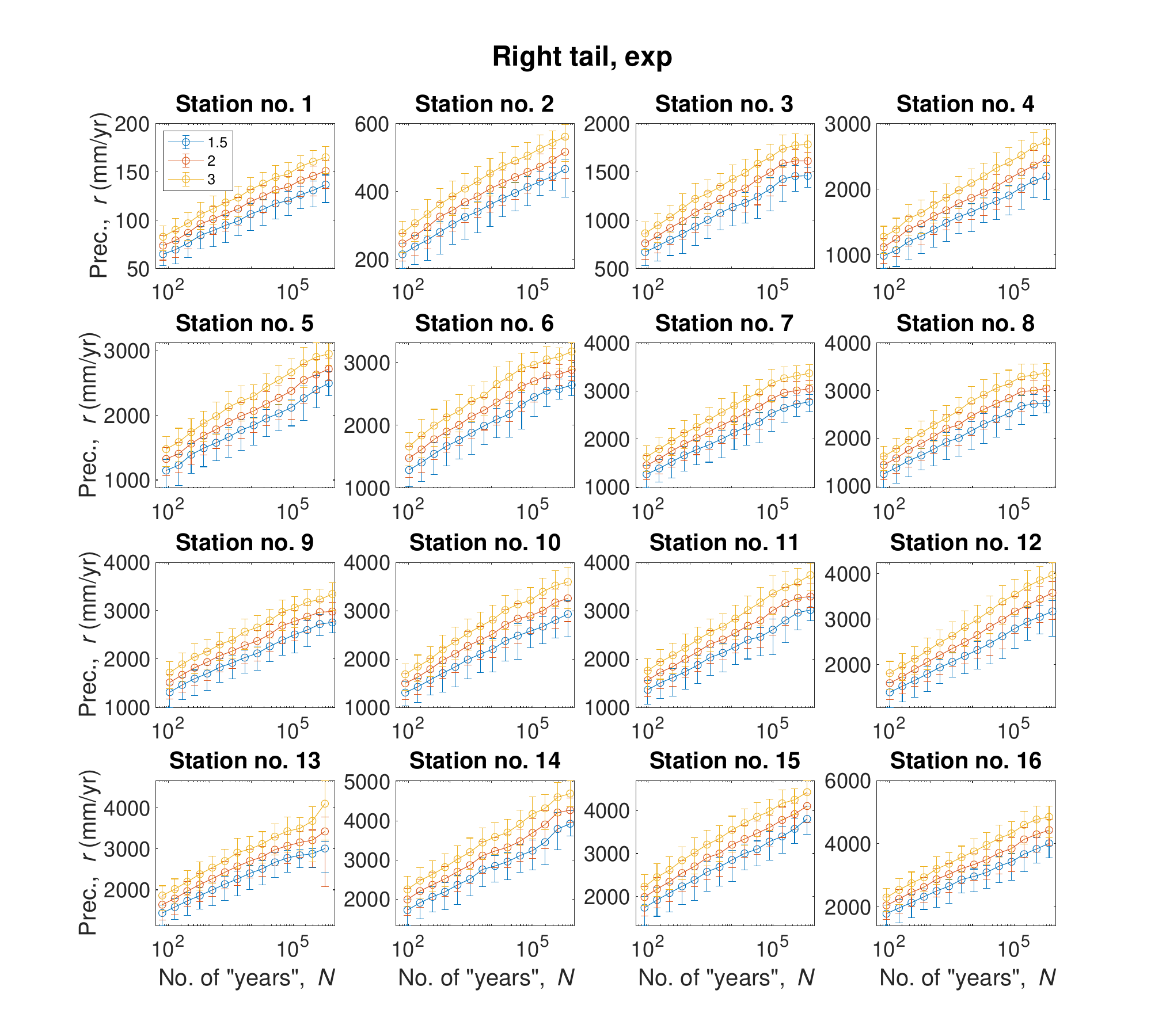}
\caption{Same as Fig.~\ref{fig_predicted_precipitation_right_tail_gamma_sort} for exponential distribution PDF model.  
}
\label{fig_predicted_precipitation_right_tail_exp_sort}
\end{figure*}

\begin{figure*}[t]
\includegraphics[width=1\linewidth]{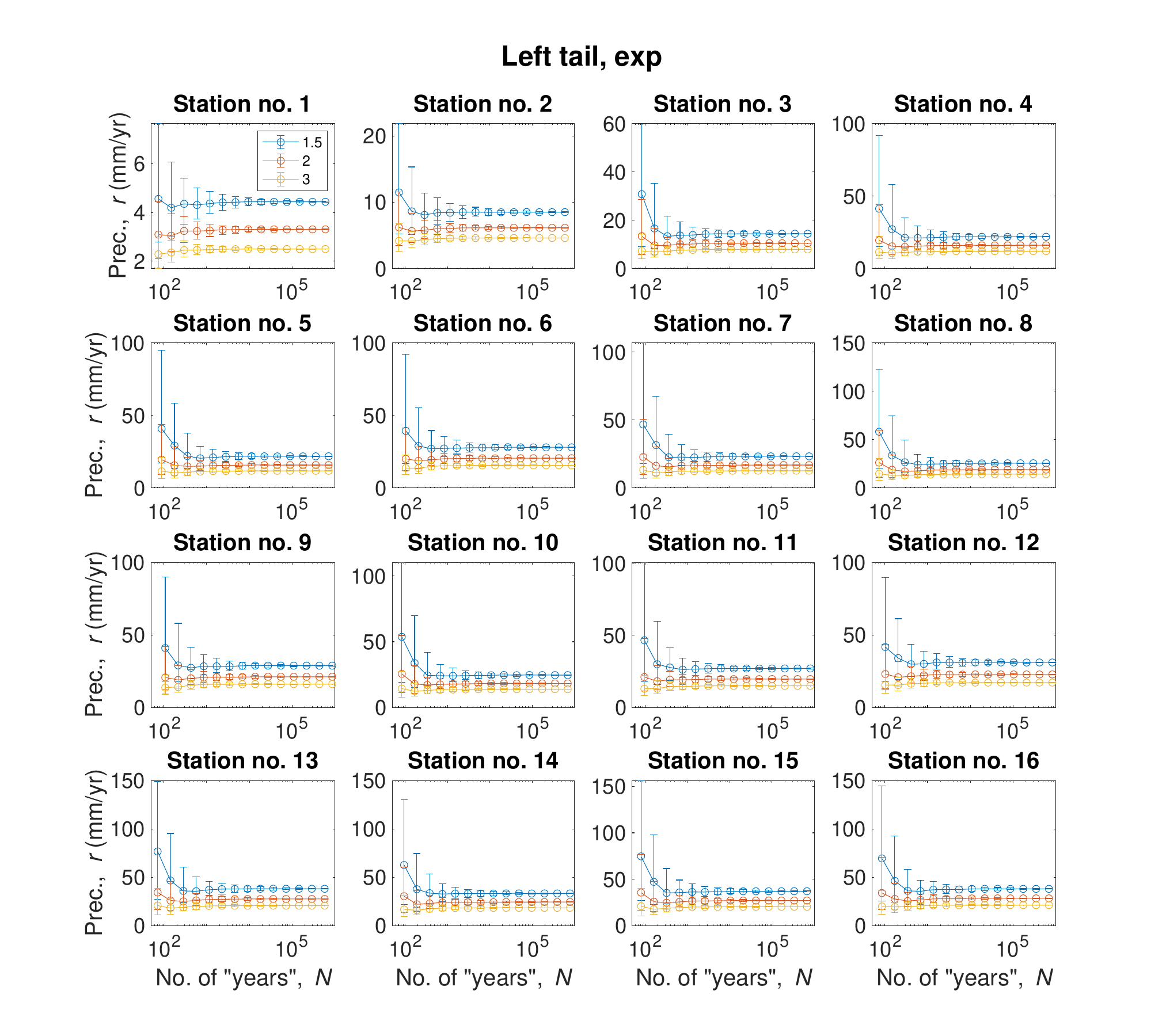}
\caption{Same as Fig.~\ref{fig_predicted_precipitation_left_tail_gamma_sort} for exponential distribution PDF model.
}
\label{fig_predicted_precipitation_left_tail_exp_sort}
\end{figure*}

%
%

\clearpage



\bibliography{all}

\end{document}